\documentclass[10pt,prl,aps,twocolumn,superscriptaddress,preprintnumbers]{revtex4-1}

\usepackage[utf8]{inputenc}
\usepackage{epigraph}
\usepackage{amsmath}
\usepackage{wrapfig}
\usepackage{epsfig}
\usepackage{amssymb}
\usepackage{graphicx}
\usepackage{slashed}
\usepackage{xcolor}
\usepackage{comment}
\usepackage{natbib}
\usepackage{enumitem}
\usepackage{tikz-cd,tikz}
\usepackage{float}
\usepackage{array,adjustbox,booktabs}
\usepackage{subcaption}
\usepackage{amsfonts}
\usepackage{physics}
\usepackage{hyperref}
\usepackage{mdframed}

\usepackage[scr=esstix,  
					]{mathalpha}

\usepackage{xcolor}
\usepackage{colortbl}
\definecolor{dark-red}{rgb}{0.80,0.12,0.12} 
\definecolor{dark-blue}{rgb}{0,0.15,0.85} 

\hypersetup{
  linktocpage,
  colorlinks  = true, 
  urlcolor    = magenta, 
  linkcolor   = dark-blue, 
  citecolor   = purple 
}

\usepackage[a4paper]{geometry}
\DeclareMathAlphabet\mathbfcal{OMS}{cmsy}{b}{n}

\usetikzlibrary{decorations.markings}
\tikzset{middlearrow/.style={
		decoration={markings,
			mark= at position 0.5 with {\arrow{#1}} ,
		},
		postaction={decorate}
	}
}

\def\be{\begin{equation}}
\def\ee{\end{equation}}

\def\S{{\bold s}}
\def\T{{\bold t}}
\def\U{{\bold u}}
\def\z{{\bold z}}

\def\tn{  {s}  }
\def\gv{  {\sigma}  }

\def\intsum{\mathop{\sum \kern-1.2em \int}\limits_{\check{s}}d \hat{s}}

\def\intsumdeltaS{\mathop{\sum \kern-1.2em \int}\limits_{\sum_j\!\check{s}_{ij}=\,p_i-\delta_{\rm S}}\kern-1.5em{d} \hat{s}}
\def\intsumuno{\mathop{\sum \kern-1.2em \int}\limits_{\sum_j\!\check{s}_{ij}=\,p_i-1}\kern-1.5em{d} \hat{s}}
\def\intsumdue{\mathop{\sum \kern-1.2em \int}\limits_{\sum_j\!\check{s}_{ij}=\,p_i-2}\kern-1.5em{d} \hat{s}}
\def\intsumtre{\mathop{\sum \kern-1.2em \int}\limits_{\sum_j\!\check{s}_{ij}=\,p_i-3}\kern-1.5em{d} \hat{s}}
\def\intsumAdS{\mathop{\sum \kern-1.2em \int}\limits_{\sum_j\!\check{s}_{ij}=\,p_i-\delta_{\rm AdS}}\kern-2.2em{d} \hat{s}}

\def\p{\underline{p}}

\newcommand{\nocontentsline}[3]{}
\let\origcontentsline\addcontentsline
\newcommand\stoptoc{\let\addcontentsline\nocontentsline}
\newcommand\resumetoc{\let\addcontentsline\origcontentsline}

\begin{document}

\title{Simplicity of Mellin amplitudes for AdS$_3 \times$S$^3$}
\author{F.~Aprile}
\email{faprile@ucm.es}
\affiliation{Departamento de F\'isica Te\'orica \& IPARCOS, Facultad de Ciencias F\'isicas, Universidad Complutense, 28040 Madrid}
\author{C.~Behan}
\email{connorbehan@ictp-saifr.org}
\affiliation{ICTP South American Institute for Fundamental Research, Instituto de F\'isica Te\'orica  UNESP Rua Dr. Bento Teobaldo Ferraz 271, 01140-070, São Paulo, SP, Brazil}
\author{R.~S.~Pitombo}
\email{rs.pitombo@unesp.br}
\affiliation{ICTP South American Institute for Fundamental Research, Instituto de F\'isica Te\'orica  UNESP Rua Dr. Bento Teobaldo Ferraz 271, 01140-070, São Paulo, SP, Brazil}
\affiliation{Mathematical Institute, University of Oxford, Andrew Wiles Building, Radcliffe Observatory Quarter, Woodstock Road, Oxford, OX2 6GG, U.K}
\author{M.~Santagata}
\email{michelesa@ntu.edu.tw}
\affiliation{Department of Physics, National Taiwan University, Taipei 10617, Taiwan}
\begin{abstract}
\noindent 
The spectrum of half-BPS single-particle operators of the D1-D5 system in the supergravity regime is dual to  
the spectrum of Kaluza-Klein modes of tensor and graviton multiplets in AdS$_3\times$S$^3$.
We present \emph{simple} formulae for all four-point tree-level correlators of scalar half-BPS  primary operators in the spectrum, 
and in particular, we bootstrap the four-point correlator of the graviton multiplet for \emph{all} KK levels.  
A key insight of our approach is the use of generalized Mellin space, 
and a decomposition of the correlators into distinct subsectors each 
one characterized  by different on-shell constraints  among the Mellin variables. 
The resulting Mellin amplitudes involve a small set of building blocks with 
manifest properties under a six-dimensional symmetry.
\end{abstract}

\maketitle

\stoptoc

\section{Introduction}
Pulling the \emph{strings} of gravity in AdS/CFT depends greatly 
on the ability to compute boundary correlation functions, especially in those superconformal 
field theories (SCFT) that have a weakly coupled gravitational dual. 
A paradigmatic example is that of $\mathcal{N}=4$ super Yang-Mills (SYM):  
in the regime where the theory is dual to supergravity on AdS$_5\times$S$^5$ 
all four-point correlators that describe tree-level scattering of Kaluza-Klein (KK) modes of the type IIB graviton 
are known in closed form thanks to the bootstrap approach pioneered in \cite{Rastelli:2016nze,Rastelli:2017}.
In this letter we extend these results to the D1-D5 system {dual to} AdS$_3\times$S$^3$ in the supergravity regime, 
which is another paradigmatic example of AdS/CFT considered in the seminal paper \cite{Maldacena:1997re}. 
Following the bootstrap approach, 
and previous work done in \cite{Giusto:2018ovt,Giusto:2019pxc,Rastelli:2019gtj, Giusto:2020neo,Aprile:2021mvq,Wen:2021lio,Behan:2024srh},
our main purpose is to determine \emph{all} tree-level half-BPS scalar four-point correlators in this theory.

The most novel and relevant aspect of the D1-D5 system is the fact that 
the spectrum of half-BPS primary operators consists of two distinct  
multiplets: the tensor multiplets and the graviton multiplets.
These are coupled to each other, and as a result, there are three types 
of four-point correlators that one could consider. 
The earliest studies  focused  on KK correlators of tensor multiplets 
showing that these can be organized into a single correlator of 6D 
scalar fields  \cite{Giusto:2018ovt,Giusto:2019pxc,Rastelli:2019gtj}, closely 
resembling the structure that \cite{Caron-Huot:2018kta} found in AdS$_5\times$S$^5$.
However, the 6D origin of the scalar primary operators
in the graviton multiplets is a self-dual three-form  \cite{Giusto:2020neo}, rather 
than a scalar field. Dealing with the polarisations of this three-form 
introduces several challenges. 
For example,  
the mixed tensor-graviton correlators can be given in a 6D 
conformally covariant way \cite{Giusto:2020neo},  but the explicit expressions 
\cite{Wen:2021lio} for all KK levels are unwieldy, and therefore a similar 
approach has not been attempted for the four gravitons correlators. 

The rigorous bootstrap approach of \cite{Behan:2024srh} yields 
explicit results for infinite families of graviton correlators, that could not be obtained 
otherwise, and therefore solves the problem of generating data for the graviton correlators. 
However, the bootstrap approach does not automatically give a formula with 
explicit dependence on the KK level,  since the latter have to be specified on a 
case-by-case basis, and a bottleneck of this approach is the
fact that the number of Witten diagrams needed to build the ansatz for the correlator 
grows fast with the KK levels, making the computation unfeasible.
Thus in order to organize the output, and prove the result for all KK levels, 
there is really the need for an additional principle.
We shall see that this principle is naturally found when the correlators 
are written in generalised AdS$_{3}\times$S$^{3}$ Mellin space, 
a version of Mellin space where both 
AdS and S are on the same footing. 

First of all, Mellin space has long been recognized to be a powerful tool for studying 
holographic correlators \cite{Mack:2009,Penedones:2010}, owing to the fact that, 
under a properly defined Mellin transform, a holographic correlator is dual
to an amplitude whose structure makes the flat-space limit manifest \cite{Penedones:2010}. 
Following  \cite{Aprile:2020luw,Abl:2020dbx,Aprile:2020mus,Aprile:2021mvq}, we will consider a 
generalization of Mellin space that includes a discrete Mellin transform on the R-symmetry 
coordinates, leading to the notion of an AdS$_{3}\times$S$^{3}$ 
Mellin amplitude. In this setting the flat-space limit can be uplifted to 6D,
by the same argument as in \cite{Aprile:2020luw}.
The AdS$_3\times$S$^3$ Mellin amplitude, which generically 
depends on all Mellin variables and the KK levels,
{therefore encodes} a 6D flat-space  amplitude.
Then, a symmetry enhancement  takes place in all cases 
in which the  AdS$_{3}\times$S$^{3}$ Mellin amplitude  only depends on those combinations of variables that 
reorganize into 6D Mandelstam invariants. This special feature of the amplitude trivializes
the dependence on the KK levels. Crucially, even when the symmetry enhancement is approximate, 
it still serves as an organizing principle  for an otherwise increasingly complex bottom-up ansatz of Witten diagrams. Making the 6D symmetry manifest  is the basic idea that we shall 
exploit to organize the scalar correlators of the D1-D5 system.
The result for the four tensor correlators \cite{Giusto:2018ovt,Giusto:2019pxc,Rastelli:2019gtj} 
was given in \cite{Aprile:2021mvq}. Here we will consider the cases of the
mixed tensor-graviton correlators and the four graviton correlators, since so far, in both these cases, the existing formalism to write down these correlators does not 
make both the symmetries and the KK level dynamics manifest.

Our main result in this letter is to demonstrate the \emph{simplicity} of the three scalar types of single particle 
correlators of the D1-D5 SCFT$_2$ in AdS$_3\times$S$^3$, by providing explicit and compact expressions 
for \emph{all} KK levels, in particular, the complete four-point graviton correlators.

\vspace{-.25cm}
\section{Four-point correlators}

In the D1-D5 SCFT$_2$ dual to type IIB string theory on AdS$_3\,\times\,$S$^3\!\times M$, 
we will denote the primary operators dual to gravitons by $\gv_p$ where $p$ is the KK level. 
The number of tensor multiplets arising from the reduction of R fields is equal to the Hodge 
number $h^{1,1}$ of $M$. There is also one from NS fields, which is on the same footing in 
supergravity (SUGRA), so we will denote these primary operators by
{$\tn_p^i$ where $i$}
is a fundamental $SO(h^{1,1} + 1)$ index.
Note that $M=T^4$ has $h^{1,1} = 4$ and leads to (2, 2) SUGRA 
while $M=K3$ has $h^{1,1} = 20$ and leads to (2, 0) SUGRA.
This letter will be specific to $K3$, closely following the 
 approach of  \cite{Rastelli:2019gtj,Behan:2024srh} that made use of the 
$(2, 0)$ three-point couplings from \cite{Arutyunov:2000}. However, we expect that similar 
results hold for $T^4$ as well. In particular, the extra supersymmetry in $(2, 2)$ forces 
one of the tensor multiplets \cite{footnote1}
to have the same 
tree-level amplitudes of the $(2, 0)$ theory \cite{Heydeman:2018}, and therefore, as discussed in \cite{Giusto:2019pxc}, 
this leads to agreement of four-point functions when all $SO(h^{1,1} + 1)$ indices are the same.

The correlators we will study are 
\begin{equation}
\label{theoldones}
\langle \tn_{p_1}^{i_1} \tn_{p_2}^{i_2} \tn_{p_3}^{i_3} \tn_{p_4}^{i_4}\rangle\qquad
\langle \tn_{p_1}^{i_1} \tn^{i_2}_{p_2} \gv_{p_3} \gv_{p_4}\rangle
\end{equation}
and 
\begin{equation}\label{thenewone}
\langle \gv_{p_1} \gv_{p_2} \gv_{p_3} \gv_{p_4}\rangle.
\end{equation}
An operator is inserted at spacetime coordinates, ${x}_i,\bar{x}_i$, and is specified by R-symmetry 
SU(2)$_L\times$SU(2)$_R$ spinors $\big(\,^{\,1}_{y_i}\big),\big(\,^{\,1}_{\bar{y}_i}\big)$. 
Cross ratios $x,\bar{x}$, $y,\bar{y}$, are defined by
\begin{equation}
x=\frac{x_{12} x_{34}}{x_{13} x_{24}}\qquad;\qquad y = \frac{y_{12} y_{34}}{y_{13} y_{24}},
\end{equation}
with $x_{ij}=x_i-x_j$, $y_{ij}=y_i-y_j$, and $\bar{x}$, $\bar{y}$ defined similarly.

Superconformal symmetry implies that correlators with external half-BPS scalar fields $\Phi$, 
such as \eqref{theoldones} and \eqref{thenewone}, can be written in the form \cite{Behan:2021,Aprile:2021mvq}, 
\begin{equation}\label{WardIdentity1}
\!\!\langle \Phi_1 \Phi_2 \Phi_3 \Phi_4 \rangle = k+ (x-y){f} +(\bar{x}-\bar{y})\bar{{f}}+ (x-y)(\bar{x}-\bar{y}) {\cal H}
\end{equation}
where $k$ is constant in the cross ratios, $f,\bar{f}$ are single-variable functions 
determined by protected data, for example through the chiral algebra 
twist discussed in \cite{Rastelli:2019gtj}, and the reduced correlator ${\cal H}$ is a 
two-variable function that splits into a plus and a minus sector, 
\begin{align}
\label{WardIdentity2}
&
{\cal H}=  x_{13}^2 x_{24}^2 y_{13}^2 y_{24}^2 \, \Big[ \mathcal{H}^{+}(x^2_{ij},y^2_{ij}) + \\
\label{WardIdentity3}
&\rule{1cm}{0pt}-(x {-} \bar{x}) (y{-} \bar{y})\, x_{13}^2 x_{24}^2 y_{13}^2 y_{24}^2\, \mathcal{H}^-(x^2_{ij},y^2_{ij}) \Big]\,,
\end{align}
with $x_{ij}^2 \equiv x_{ij}\bar{x}_{ij}$ and $y^2_{ij}\equiv y_{ij}\bar{y}_{ij}$.
The reduced correlators ${\cal H}^{\pm}$ are symmetric functions of both 
$x,\bar{x}$ and $y,\bar{y}$. However, unlike higher-dimensional CFTs such 
as $\mathcal{N}=4$ SYM in 4D, only the \emph{simultaneous} exchange 
of  $x \leftrightarrow \bar{x}$, $y \leftrightarrow \bar{y}$ leaves the full 
correlator invariant. See again \eqref{WardIdentity1}, \eqref{WardIdentity2} 
and \eqref{WardIdentity3}. 

In the supergravity regime, the leading $O(1)$ contributions -- in the large central 
charge $c$ expansion -- are disconnected two-point functions, when they exist. 
These can be rearranged into the form dictated by superconformal symmetry, 
thus giving a contribution to $k,f,\bar{f}$ and ${\cal H}^{\pm}$.  
In general, free-type contributions have a vanishing Mellin transform and we 
will not consider them further. 

{\bf Generalized Mellin space.} We are interested in the tree-level connected 
$O(1/c)$ contribution to $\mathcal{H}^{\pm}$, and we will present results using the Mellin transform,
\begin{equation}\label{iniMellin}
{\cal H}^{\pm}_{p_1,p_2,p_3,p_4} = \intsum\, {\cal M}^{\pm}( \hat{s}_{ij},\check{s}_{ij};p_1,p_2,p_3,p_4)
\end{equation}
with the AdS$_3\times$S$^3$ integration ``measure"  defined by
\begin{equation}\label{eqini}
\intsum \, 
\equiv  \sum_{\tilde{s},\tilde{t}}  \oint_{-i\infty}^{+i\infty}\!\!\frac{dsdt}{(2\pi i)^2} \prod_{i<j}   
{ x_{ij}^{2 \hat{s}_{ij}}} {  y_{ij}^{2 \check{s}_{ij}} } \frac{\Gamma[-\hat{s}_{ij}]}{\Gamma[\check{s}_{ij}{+}1] }\,.
\end{equation} 
The Mellin variables are symmetric $\hat{s}_{ij}=\hat{s}_{ji}$, $\check{s}_{ij}=\check{s}_{ji}$, 
off-diagonal, $\hat{s}_{ii}=\check{s}_{ii}=0$,  and constrained by ``on-shell" relations 
that depend on $p_i$ and have the form
\begin{equation}\label{onshellrel}
\sum_{j=1}^4\check{s}_{ij} = p_i - \delta_{\text{S}}\,,\ \qquad\   \sum_{j=1}^4\hat{s}_{ij} =  -p_i -\delta_{\text{AdS}} \,,
\end{equation}
where $\delta_{\text{S}},\delta_{\text{AdS}}\in\mathbb{N}$.
Solving the constraints gives four independent variables, conjugate to the cross ratios. 
We shall call  $s$ and $t$  the independent variables in the $\hat{s}_{ij}$, and
 $\tilde{s}$ and $\tilde{t}$ those in the $\check{s}_{ij}$. See \eqref{eqini}.

We will always work with fixed values of $\delta_{\text{AdS}}$,
\begin{equation}
\delta_{\text{AdS}}=1\ \,\textrm{for}\ \,{\cal H}^+\,,\ \qquad\  \delta_{\text{AdS}}=2\ \,\textrm{for}\ \,{\cal H}^-\,.
\end{equation}
Regarding $\delta_{\text{S}}$, we will fix it a posteriori 
and add this information to the symbol defined in \eqref{eqini}.
This is a crucial novelty of our formalism, and is
motivated by the idea of simplifying the dependence on the R-symmetry variables in the Mellin amplitude.
To see how, let us recall that ${\cal H}^{\pm}$ are by construction polynomials in the $y_{ij}^2$. 
This polynomiality shows up in the measure 
as a consequence of the fact that the $\Gamma[\check{s}_{ij}+1]^{-1}$ vanish unless $\check{s}_{ij}\ge 0$ (which also
shows that the sum \eqref{eqini} can be kept unbounded). Then, because $\check{s}_{ij}\ge 0$, 
the corresponding on-shell constraints in \eqref{onshellrel} imply that unless $p_i\ge \delta_{\text{S}}$ there 
is no contribution to the Mellin amplitude. Thus, the value of $\delta_{\rm S}$ can 
be used as a knob to stratify the (numerator of the) Mellin amplitude in the
R-symmetry variables. Note however that superconformal symmetry, see \eqref{WardIdentity1}-\eqref{WardIdentity3},
gives from the start that $\delta_{\text{S}}\ge1$ 
for ${\cal H}^+$ and $\delta_{\text{S}}\ge2$ for ${\cal H}^-$. We shall see that 
$\delta_{\text{S}}=1,2,3$ are the values that we need.

{\bf Bold-face variables and 6D symmetry.} 
A set of variables that will feature a special role are 
the bold-face variables defined via
\begin{equation}
{\bf s}_{ij}= \check{s}_{ij}+\hat{s}_{ij}\qquad;\qquad \sum_j{\bf s}_{ij} = -\delta
\end{equation}
where  $\delta\equiv \delta_{\text{AdS}}+ \delta_{\text{S}}$.
Note that since the constraints do not depend on the $p_i$, 
 there is really only one bold-variable in each channel, as for scattering amplitudes, 
\begin{align}
\label{boldface}
  {\bf{s}}   \equiv  {\bf s}_{12}={\bf s}_{34},\quad 
 {\bf{t}}  \equiv {\bf s}_{14}={\bf s}_{23}\,,\quad
 {\bf{u}} & \equiv {\bf s}_{13}={\bf s}_{24},
 \end{align}
with ${\bf{s}} +{\bf{t}} +{\bf{u}}=-\delta$. 
A corresponding 6D Mellin transform in these variables is defined by using the measure
\begin{equation}\label{6Dmeasure}
\begin{tikzpicture}
\draw (0,0) node[scale=.9] {$
\displaystyle\!\!\!\oint\!{d}\pmb\mu_{\delta}\equiv  \oint\!\!\frac{d\S d\T}{(2\pi i)^2} ( \z^2_{12} \z^2_{34} )^{\S}  ( \z^2_{14} \z^2_{23} )^{\T}  ( \z^2_{13} \z^2_{24} )^{\!{\U}} \Gamma[-\S]^2\Gamma[-\T]^2 \Gamma[-{\U}]^2$};
\end{tikzpicture}
\end{equation}
where ${\z}_{ij}^2=x_{ij}^2-  y_{ij}^2$ is now a 6D distance.

The AdS$_3\times$S$^3$ measure \eqref{eqini} and the 6D measure \eqref{6Dmeasure} are closely related.
In particular, for any AdS$_3\times$S$^3$ Mellin amplitude that
depends only on bold-face variables, the KK reduction from 6D to AdS$_3\times$S$^3$ is implemented by the following identity, 

\begin{equation}\label{resummation}
\sum_{p_i\, \ge\, \delta_{\rm S}}
\intsum \, \mathcal{M}({\bf s},{\bf t}) =  \, \displaystyle\!\!\!\oint\!{d}\pmb\mu_{\delta}\,  \mathcal{M}({\bf s},{\bf t}) \,,
\end{equation}
which holds as an expansion in the $y_{ij}^2$. This shows that
if an AdS$_3\times$S$^3$ Mellin amplitude
depends only on bold-face variables, then it corresponds to a 6D conformal integral. 

Before presenting results for $\langle \gv_{p_1} \gv_{p_2} \gv_{p_3} \gv_{p_4}\rangle$,
our first goal in the next sections will be to recast
the known correlators \cite{Giusto:2019pxc,Rastelli:2019gtj,Giusto:2020neo,Wen:2021lio}
as AdS$_3\times$S$^3$ Mellin amplitudes.
The result for $\langle\tn_{p_1}^{i_1} \tn_{p_2}^{i_2} \tn_{p_3}^{i_3} \tn_{p_4}^{i_4}\rangle$ was given in \cite{Aprile:2021mvq}, and
since its 6D origin is a scalar amplitude the rewriting follows immediately from the identity \eqref{resummation}. 
This story is analogous to AdS$_5\times$S$^5$  \cite{Caron-Huot:2018kta}.
Instead, the rewriting for the tensor–graviton correlators $\langle \tn_{p_1}^{i_1} \tn^{i_2}_{p_2} \gv_{p_3} \gv_{p_4}\rangle$ is new.
For the latter we expect interesting modifications, in addition to \eqref{resummation},
because the 6D origin is not a scalar amplitude.

\vspace{-.25cm}
\section{Tensor correlator}

The four-tensor correlators are the simplest. 
For the plus-type correlators \cite{Aprile:2021mvq} found that
\begin{align}\label{4tensors}
 \mathcal{H}_{\p}^{i_1i_2i_3i_4;+} =	\intsumuno \, \mathcal{M}^{i_1i_2i_3i_4} \,,
\end{align}
with the Mellin amplitude
\begin{equation}
\label{fourtensor_mellin}
\mathcal{M}_{}^{i_1i_2i_3i_4} = -  \frac{\delta^{i_1 i_2}{\delta^{i_3 i_4}}}{{\bf{s}} +1}- \frac{\delta^{i_1 i_4}{\delta^{i_2 i_3}}}{{\bf{t}}+1}- \frac{\delta^{i_1 i_3}{\delta^{i_2 i_4}}}{{\bf{u}} +1}\,,
\end{equation}
where ${\bf{s}} +{\bf{t}} +{\bf{u}}=-2$.
See supplemental material sect.~I for the normalization we use.
The 6D symmetry in 
\eqref{4tensors}-\eqref{fourtensor_mellin} is completely manifest, since the amplitude depends only 
on bold variables. Using identity \eqref{resummation} we find 
\begin{equation}\label{identity_tensor}
\sum_{p_i\ge 1} 
\, \mathcal{H}_{\p}^{i_1i_2i_3i_4;+} = \oint\!{d}\pmb\mu_{\delta=2}\, \mathcal{M}^{i_1i_2i_3i_4} \,.
\end{equation}
This shows that the KK tower of correlators in \eqref{4tensors} originates from a  6D
correlator of dimension $\delta=2$ scalar fields, as this is precisely the value appearing in the
{on-shell constraints for the bold-face variables}.

The ${\cal H}^-$ correlators vanish at this order. This can be shown both from 
the bootstrap \cite{Rastelli:2019gtj, Behan:2024srh} and from the HHLL limit of \cite{Giusto:2019pxc}. From 6D it is 
also clear that it is not possible to construct a tree-level scalar function of the ${\bold z}^2_{ij}$ that gives 
$(x-\bar{x})(y-\bar{y})$. Nevertheless, the disconnected contributions to the minus sector 
are non-trivial, and similarly loop contributions \cite{FAMS}.

\vspace{-.25cm}
\section{Tensor-graviton correlator}

The mixed tensor-graviton correlator was obtained in 6D notation in \cite{Giusto:2020neo}. 
The scalar primaries in the graviton multiplets are KK modes of a self-dual 
three-form with polarizations projected appropriately.
The bootstrap approach \cite{Behan:2024srh}  gives results in agreement with the 6D derivation,
thus providing conclusive evidence for the higher-dimensional symmetry.

Since the flavor symmetry is fully diagonal in this case, we will consider 
\begin{equation}
\mathcal{H}_{\p}^{i_1i_2} =  \delta^{i_1i_2} \mathcal{H}_{\p}\,
\end{equation}
and present results for $\mathcal{H}^{\pm}_{}$ using generalized Mellin space.
In the supplemental material sect.~III we explain 
how to go from the 6D formula of \cite{Giusto:2020neo} to our new formula.

The minus-type correlator reads
\begin{align}\label{tensgravmin}
 \mathcal{H}_{\p}^{-}\  =   	\!\! \intsumdue \  \mathcal{M}_{stu}
\end{align}
with the Mellin amplitude 
\begin{equation}
\label{tensgravMellinmin}
\mathcal{M}_{stu}= \frac{1}{({\bf{s}} +1)({\bf{t}}+1)({\bf{u}} +1)} 
\end{equation}
where ${\bf{s}}+{\bf{t}} +{\bf{u}}=-4$. Interestingly, despite the fact that the mixed 
correlator is only invariant under the exchange of points $1 \leftrightarrow 2$ or  $3 \leftrightarrow 4$, 
the Mellin amplitude in \eqref{tensgravMellinmin} exhibits full crossing symmetry as function 
of ${\bold s},{\bold t},{\bold u}$. Moreover, it is the same function that 
appears in AdS$_5 \times$S$^5$ \cite{Aprile:2020luw}. To understand why 
let us recall that the superconformal block expansion of all correlators with external 
half-BPS scalar primaries, \eqref{WardIdentity1}, when restricted to the ${\cal H}^-$ sector, is 
identical to that of ${\cal N}=4$ SYM in the long sector \cite{Aprile:2021mvq,Behan:2024srh}.
This suggests that if a resummation like \eqref{resummation} exists for the minus-type correlators, 
it is likely to involve the same functions as in ${\cal N}=4$ SYM, and indeed this is what we found
\begin{equation}\label{identity_minus}
\sum_{p_i\ge 2}  
\,\mathcal{H}_{\p}^-  = \oint\!{d}\pmb\mu_{\delta=4}\, \mathcal{M}_{stu} \,.
\end{equation} 
We emphasize that the origin of this result is six-dimensional, 
and is a consequence of the fact that the 6D origin of ${\cal H}_{\p}$ is \emph{not} a scalar amplitude.
In particular, the 6D result of \cite{Giusto:2020neo} not only generates the 
amplitude in \eqref{tensgravMellinmin}, but the full result 
$x_{13}^2 x_{24}^2 y_{13}^2 y_{24}^2(x {-} \bar{x}) (y{-} \bar{y}){\cal H}^-_{\p}$, 
including the prefactors!

{\bf A new formula for the plus-type correlator.}
These correlators appear from the start to be non-trivial because the three-form polarizations give structure to the numerator of the amplitude.  
This is evident in the way the result is derived from 6D~\cite[(4.1)]{Giusto:2020neo}.  The crucial idea is to 
organize this structure in terms of 
a convenient set of variables, %
\begin{equation}\label{naturalvar}
y_{ij}^2\qquad;\qquad \frac{ y_{ij}^2}{\z_{ij}^2}\,.
\end{equation}
Our expectation is that poles in the bold-face variables can only be 
accompanied by the $y_{ij}^2$ themselves or by the 6D ``propagators" ${ y_{ij}^2}/{\z_{ij}^2}$, 
which can be understood as operators acting on scalar Mellin amplitudes in 6D. 
As a result we  find that $\mathcal{H}_{}^{+}$ is naturally stratified over 
different R-symmetry factors, and given by
\begin{widetext}
\begin{align}\label{tensgravcorr}
\mathcal{H}_{\p}^{+}= &\!\!\!\intsumuno\,\mathscr{M}_{1s}\ \,
+\!\!\!\!\!\intsumdue\left( y_{12 }^2 y_{34}^2 \mathscr{M}_{2s} + \mathcal{Y}_{s}\mathcal{M}_{tu}+ \mathcal{Y}_{t}\mathcal{M}_{su}+ \mathcal{Y}_{u}\mathcal{M}_{st} \right) 
\ \, -\!\!\!\!\!\intsumtre\left( y_{13}^4  y_{24}^4 (y-\bar{y})^2  \mathcal{M}_{stu} \right)
\end{align} 
\end{widetext}
with $\mathcal{Y}_{u} =  (y_{13}^2  y_{24}^2{-}y_{23}^2  y_{14}^2{-}y_{12}^2  y_{34}^2)$, 
and $\mathcal{Y}_{s},\mathcal{Y}_{t}$ defined similarly, and the new generalized 
Mellin amplitudes defined as follows. First
\begin{equation}\label{tensgravplusMellin1}
\mathscr{M}_{1s} =  \frac{ 2f_{34}}{{\bf s}+1} 
\end{equation}
with 
\begin{equation}
\label{fdef}
f_{ij} \equiv \left(\!\frac{p_i+p_j-1}{2}-\check{s}_{ij}\!\right)^{\!\!2} - \frac{p_i^2+p_j^2-1}{4} \,.
\end{equation}
Then,  
\begin{align}
\label{M2stensgrav}
&
\mathscr{M}_{2s} = - \frac{2}{{\bf s}+1} \,,\\
\label{Mst}
&
\mathcal{M}_{st}= \frac{ (\check{s}_{13}+ \check{s}_{24}+1) }{({\bf s}+1)({\bf t}+1)} \,,  \\ 
\label{Msu}
&  \mathcal{M}_{su}= \frac{ (\check{s}_{14}+ \check{s}_{23}+1) }{({\bf s}+1)({\bf u}+1)}\,, \\
\label{Mtu}
 & \mathcal{M}_{tu}= \frac{ (\check{s}_{12}+ \check{s}_{34}+1) }{({\bf t}+1)({\bf u}+1)}\,,
\end{align}
where ${\bold s}+{\bold t}+{\bold u}=-3$.
Finally, $\mathcal{M}_{stu}$ is the same as in \eqref{tensgravMellinmin} with the same on-shell constraints, ${\bold s}+{\bold t}+{\bold u}=-4$. Note that 
$-y_{13}^4  y_{24}^4 (y-\bar{y})^2=$ 
${\cal Y}_s{\cal Y}_t+{\cal Y}_s{\cal Y}_u+{\cal Y}_t{\cal Y}_u$.

Our formula \eqref{tensgravcorr} is non-trivial, and what is crucial is that
the $\check{s}$-type Mellin variables satisfy different on-shell constraints 
in correspondence of the different R-symmetry structures. 
Before moving on to the four-graviton correlator, we wish to emphasise some features 
of our result. First, much like the minus sector, the part containing simultaneous poles, 
namely $\mathcal{M}_{stu}$ and $({\cal Y}_u\mathcal{M}_{st}+\emph{crossing})$, 
are fully crossing symmetric.  Another interesting property is that the correlator 
automatically vanishes when $p_3=1$ or $p_4=1$, in agreement with the fact that the 
KK tower associated to the graviton starts from $p_i=2$.  When $p_1 = p_2 = 1$, the amplitude 
becomes proportional to $\frac{1}{{\bf s}{+}1}$, thus taking the same form as 
the $s$-channel contribution to the four-tensor amplitude, as pointed out in \cite{Behan:2024srh}.
Finally, the sub-amplitude $\mathscr{M}_{1s}$ explicitly breaks a certain $\mathbb{Z}_2$ symmetry.
We will come back to this later on.

\vspace{-.1cm}
\section{Graviton correlator}

Let us now consider the graviton correlators, for which the results here are \emph{new}, 
and so far there is no 6D derivation. Inspired by the tensor-graviton results and guided 
by explicit results obtained using the bootstrap \cite{Behan:2024srh}, 
which we recall in the supplemental material sect.~II, 
we were able to derive a closed-form expression for all KK levels.  As we are going to see 
in a moment, the structure is remarkably similar to the tensor-graviton correlator.

{ \bf Minus-type correlator.}
The minus sector is 
\begin{align}\label{gravmin}
 \mathcal{H}_{\p}^{-} =\ d_{\p}\intsumdue  \ \ \mathcal{M}_{stu}
\end{align}
where $\mathcal{M}_{stu}$ is exactly the same as \eqref{tensgravMellinmin}, 
as suggested by the same argument used for the mixed tensor-graviton correlator,
and the new normalisation factor is given by
\begin{equation}\label{dunderp}
d_{\p} = \frac{1}{2} \sum_i (p_i^2-1)\,. 
\end{equation}

{ \bf Plus-type correlator.}
The plus sector exhibits the same structural decomposition as the tensor-graviton correlator. 
Once again, the key observation is that the Mellin amplitude naturally splits  into distinct sub-amplitudes, each one accompanied by a different prefactor:
\begin{widetext}
\begin{align}\label{gravmelltotplus}
\mathcal{H}_{\p}^{+}= \!\!\!\intsumuno\,\left( \mathcal{M}_{1s}+ \textrm{\emph{crossing}}\right)\ \,
&+\!\!\!\!\!\intsumdue\left( y_{12 }^2 y_{34}^2 \mathcal{M}_{2s} + \textrm{\emph{crossing}}  + d_{\p}\, \mathcal{Y}_{u}\mathcal{M}_{st} + \textrm{\emph{crossing}} \right) \\[.2cm]
&\label{gravmelltotplus2}
 +\!\!\!\!\!\intsumtre\left( y_{12 }^4 y_{34}^4 \mathcal{M}_{3s} + \textrm{\emph{crossing}} +  d_{\p}\, y_{13}^4  y_{24}^4 (y-\bar{y})^2  \mathcal{M}_{stu}\right)
\end{align} 
\end{widetext}
where ${\cal M}_{st},{\cal M}_{su},{\cal M}_{tu}$ and ${\cal M}_{stu}$ are the same 
as in \eqref{tensgravMellinmin}, \eqref{Mst}-\eqref{Mtu},  and the new generalized Mellin amplitudes are:
\begin{equation}\label{M1gg}
 \mathcal{M}_{1s} = -4 \frac{ f_{12}f_{34}}{{\bf s}+1} 
\end{equation}
where $f_{ij}$ was defined already in \eqref{fdef}, then 
\begin{equation}\label{M2sgg}
\mathcal{M}_{2s} =  4  \frac{f_{12} + f_{34} +\check{s}_{12}+\check{s}_{34}+2-\frac{1}{2}\sum_i p_i}{{\bf s}+1}\,,
\end{equation}
and finally
\begin{equation}\label{M3sgg}
\mathcal{M}_{3s} = - \frac{4}{{\bf s}+1}\,.
\end{equation}
As before, the on-shell relations are different depending on the sub-amplitude. 
Note again that the correlators automatically vanish unless $p_i\ge 2$. 

\vspace{-.1cm}
\section{Properties of the Amplitudes}

Taking a closer look at our results we see that all three types of amplitudes 
have striking features.  Firstly, all poles are given in bold-face variables. 
Secondly, all amplitudes depend on a small set of 
building blocks:  the sub-amplitudes contributing to simultaneous 
poles,  ${\cal M}_{stu}$ and $({\cal Y}_u {\cal M}_{st} +\text{\emph{crossing}})$, 
that appear in both the mixed tensor-graviton correlators \eqref{tensgravcorr}
and the four graviton correlators \eqref{gravmelltotplus}-\eqref{gravmelltotplus2},
with the only difference given by the numerical factor $d_{\p}$ in \eqref{dunderp};
and  the amplitudes contributing to single poles.
{The latter}, $\mathscr{M}_{1s}$, $\mathscr{M}_{2s}$, and ${\cal M}_{1a}$, 
${\cal M}_{2a}$, ${\cal M}_{3a}$, $a=s,t,u$, {despite depending}
on the type of correlator, still follow a simple pattern.

The fact that all poles appear in bold-face variables has a simple explanation from 6D: 
it follows from the structure of the propagators, see e.g.~\emph{identity 1, 2, 3} of sect.~III in the supplemental material. 
From AdS$_3\times$S$^3$, a similar conclusion 
follows from considering the large $p$ limit \cite{Aprile:2020luw}, where $p_i/n$ is fixed with $n\rightarrow\infty$. 
In this limit the measure  \eqref{eqini} has a saddle point where   $\hat{s}_{ij}/n$ and $\check{s}_{ij}/n$ are 
finite and recombine to become the 6D Mandelstam invariants 
of four 
(massless) geodesics meeting at a point. Thus ${\bf{s}}_{\rm flat} +{\bf{t}}_{\rm flat} +{\bf{u}}_{\rm flat}=0$,
with the AdS$_3\times$S$^3$ Mellin amplitude encoding a 6D flat-space scattering amplitude.
Now, if ${\cal M}^{}( \hat{s}_{ij},\check{s}_{ij};\p)$
has a tree-level pole, say in the $s$-channel, this must become ${\bf s}_{\rm flat}=0$ 
in the large $p$ limit. However, ${\bold s}=\emph{constant}$ is the only possible deformation.
Same for the $t$ and $u$ channel poles. Besides this,
$\mathcal{M}_{}^{i_1i_2i_3i_4}$ is a scalar amplitude in 6D, therefore the large $p$ limit fixes it
completely at tree level. Instead, both $\langle \tn_{p_1}^{i_1} \tn^{i_2}_{p_2} \gv_{p_3} \gv_{p_4}\rangle$ in \eqref{tensgravcorr}
and $\langle \gv_{p_1}\gv_{p_2} \gv_{p_3} \gv_{p_4}\rangle$ in
 \eqref{gravmelltotplus}-\eqref{gravmelltotplus2}
have more structure, and various sub-amplitudes have non trivial numerators.
In view of the flat-space amplitudes of \cite{Heydeman:2018},
it would be very interesting to revisit the large $p$ limit in these cases, 
 and also, derive  \eqref{gravmelltotplus}-\eqref{gravmelltotplus2} from 6D.

{\bf $\mathbb{Z}_2$-symmetry.} 
Regarding the constructability of the sub-amplitudes, and their numerators, there is a 
symmetry that one might use. To begin with, note that the  
AdS$_3\times$S$^3$ measure \eqref{eqini} is  invariant under \emph{variations} of the 
charges $p_i$  such that the values $p_i+p_j$  swap in a given channel. 
Take the $s$-channel Gamma functions, for example,  after solving the on-shell 
constraints \eqref{onshellrel} one finds
\begin{equation}\label{schgamma}
\prod_{ij=12,34}\!\frac{\Gamma[\hat{s}_{ij}]}{\Gamma[1+\check{s}_{ij}]}=\prod_{ij=12,34}\!\frac{\Gamma[\frac{p_i+p_j}{2}-s]}{\Gamma[1+\frac{p_i+p_j}{2}+\tilde{s}]}\,.
\end{equation} 
If the \emph{values} of $p_1+p_2$ and $p_3+p_4$ swap, the RHS of \eqref{schgamma} is invariant, and
if at the same time the other values of $p_i+p_j$ do not change, then the measure is invariant. 
In terms of the $\check{s}_{ij}$ variables this is the $\mathbb{Z}_2$-symmetry 
\begin{equation}\label{Z2}
\check{s}_{12}\leftrightarrow \check{s}_{34},\qquad 
\hat{s}_{12}\leftrightarrow \hat{s}_{34},\qquad c_s\leftrightarrow -c_s,
\end{equation} 
where $c_s=\frac{p_1+p_2-p_3-p_4}{2}$. 
A concrete case to have in mind is $\p=\{k{+}1,k{+}1,k{-}1,k{+}1\}$ and $\p=\{k,k,k,k{+}2\}$.
Analogous considerations apply to the $t$ and $u$ channels, but
note that this symmetry is not crossing symmetry \cite{Aprile:2020mus,Aprile:2022tzr}. 
To see what combination of $p_i$ is invariant it is useful to write them in terms of 
$c_s,c_t,c_u,\Sigma=\frac{p_1+p_2+p_3+p_4}{2}$. For example, $d_{\p}$ in \eqref{dunderp} is 
quadratic and therefore invariant in all three channels. Now,  
a Mellin amplitude that depends only on bold-face variables is obviously invariant under 
the $\mathbb{Z}_2$-symmetries of the measure.
This is the case for $\mathcal{M}_{}^{i_1i_2i_3i_4}$, for $\mathscr{M}_{2s}$, ${\cal M}_{3s},{\cal M}_{3t},{\cal M}_{3u}$, and
for ${\cal M}_{stu}$,  ${\cal M}_{st},{\cal M}_{tu},{\cal M}_{su}$.
In particular,  all sub-amplitudes contributing to simultaneous poles are invariant.  
The question is more subtle in the presence of $f_{ij}$. However, ${\cal M}_{2s},{\cal M}_{2t},{\cal M}_{2u}$ 
are $\mathbb{Z}_2$-symmetric, as follows from the fact that
no degree-2 polynomial can be invariant under crossing while breaking the $\mathbb{Z}_2$-symmetry.
In contrast, a single $f_{ij}$ {or a product $f_{ij} f_{kl}$ of degree 4}, are not invariant. Consequently,
$\mathscr{M}_{1s}$ and ${\cal M}_{1s},{\cal M}_{1t},{\cal M}_{1u}$  fail to be invariant. Thus,
the sub-amplitudes contributing to simple poles with no R-symmetry factor are the only symmetry breaking terms.

\vspace{-.2cm}
\section{Summary and Outlook}

We studied tree-level four-point correlators of the chiral primary operators  
$s^i_p$ and $\sigma_p$ of the D1-D5 SCFT$_2$ in the supergravity regime, 
in the tensor and graviton multiplets, respectively.
For all three types of correlators we have shown that the AdS$_3\times$S$^3$ Mellin 
space formalism, together with the idea of reducing the Mellin amplitude to a set of 
building blocks with given on-shell constraints, dramatically simplifies their structure. 
This, and additional data from the bootstrap, has led us to the complete determination 
of $\langle \gv_{p_1} \gv_{p_2} \gv_{p_3} \gv_{p_4}\rangle$ in  \eqref{gravmelltotplus}-\eqref{gravmelltotplus2}.
Remarkably, all three types of correlators are given by simple formulae
with manifest properties under a 6D symmetry. 

While the existence of the 6D symmetry is not new, 
it was not understood how to make both symmetries and the KK level dynamics manifest.
The first purpose of our work has been to establish a formalism that allows to do so.  
In particular, our approach reduces the bootstrap problem to an over-constrained problem 
allowing a rigorous proof of all our results, thus going
beyond state-of-the-art.
%
%
%
%
%
%
%
We believe that our formalism will be useful in other studies, and we list some 
future directions: the study of correlators with vector fields that sit in the same 
6D three-form field as the $\sigma_{p}$; 
the unmixing of long double-particle operators 
and the use of unitary methods as starting point to develop a bootstrap program at loop level,
as done in higher dimensional cases \cite{Caron-Huot:2018kta,Aprile:2019rep,Huang:2021xws,Drummond:2022dxw};
the study of correlators with multi-particle 
operators built out of the $s^i_p$  \cite{Ceplak:2021wzz,Aprile:2025hlt} and/or the $\sigma_p$;
the relation between $K3$ and $T^4$ compactifications, 
motivated by a possible generalization to AdS of  
the flat-space SUSY reduction $(2,2)\rightarrow (2,0)$ in \cite{Heydeman:2018}; 
and lastly, the study of stringy corrections, generalizing the progress made in \cite{Chester:2024}.

Finally, it would be interesting to turn our attention to the 4D $\mathcal{N} = 2$ SCFT in \cite{Alday:2021odx} 
dual to AdS$_5\times $S$^3$ whose spectrum contains both gluon and gravitational fields.
The KK flavor currents from the gluons, analogous to the $\tn^i_p$ from the tensors,  sit in an 8D scalar field.
Their correlators are known in terms of generalized Mellin amplitudes at 
four-points tree-level \cite{Alday:2021odx,Drummond:2022dxd}, in the $\alpha'$ expansion \cite{Glew:2023wik}, 
and at five points \cite{Huang:2024dxr}. 
The striking relation of these amplitudes with their flat-space counterparts has
shed light on  BCJ and double-copy relations in AdS \cite{Zhou:2021gnu,Drummond:2022dxd}.
Similar results for gravitons are not yet known and are complicated 
by the fact that the $\mathcal{N} = 2$ stress tensor is typically exchanged together 
with infinitely many long multiplets \cite{Behan:2023,Chester:2023,Behan:2024}.
Instead, one might start with $SU(2)$ flavor current multiplets analogous to $\gv_p$. 
Progress computing their correlators was recently made in \cite{Chester:2025ssu}.

\vspace{-.1cm}
\section{Acknowledgments}
We thank S.~Giusto, R.~Russo, C.~Wen, X.~Zhou for discussions and comments on the draft.
C.~Behan and R.S.~Pitombo thank M.~Nocchi for collaboration on related work.
F.~Aprile is supported by
RYC2021-031627-I funded by MCIN/AEI/10.13039/501100011033 and by the NextGeneration 
EU/PRTR, and the
MSCA ``HeI" with \href{https://cordis.europa.eu/project/id/101182937}{DOI 10.3030/101182937}. 
C.~Behan and R.S.~Pitombo are supported by the São Paulo Research Foundation (FAPESP) 
through the grants 2019/24277-8, 2022/05236-1, 2023/03825-2 and 2024/08232-2.
M.~Santagata is supported by the Ministry of Science and
Technology (MOST) through the grant 110-2112-M002-006-.

\newpage
\resumetoc
\onecolumngrid
\vspace{2cm}

\setcounter{equation}{0} 
\renewcommand{\theequation}{S\arabic{equation}} 
\setcounter{secnumdepth}{2}

\begin{center}
\textbf{\large Supplemental Materials}
\end{center}

\tableofcontents

\parskip0.3cm

\section{Two-point function normalizations}\label{appendix_normalization}

In this section we clarify the two-point function normalization of the external operators, both $s_p^i$ and $\sigma_p$.
The natural normalization is the one where the external operators 
have two-point functions normalized to one. In the main letter we used a slightly different 
normalization, but the change of conventions simply amounts to adding an overall prefactor in each case.

{\bf Tensor correlators}
\begin{equation}\label{normalize_tensor}
\mathcal{H}_{\p}^{i_1i_2i_3i_4}\rightarrow \sqrt{p_1p_2p_3p_4}\ \mathcal{H}_{\p}^{i_1i_2i_3i_4}
\end{equation}

{\bf Mixed tensor-graviton correlators}
\begin{equation}\label{normalize_tensor_graviton}
\mathcal{H}_{\p}^{i_1i_2}\rightarrow \frac{\sqrt{p_1p_2p_3p_4}}{\sqrt{(p_3^2-1)(p_4^2-1)}}\ \mathcal{H}_{\p}^{i_1i_2}
\end{equation}

{\bf Graviton correlators}
\begin{equation}\label{normalize_graviton}
\mathcal{H}_{\p}^{}\rightarrow \frac{\sqrt{p_1p_2p_3p_4}}{\sqrt{\prod_{i}(p_i^2-1)}}\ \mathcal{H}_{\p}^{}
\end{equation}

In the next section, following the bootstrap approach in \cite{Behan:2024srh},  we will consider 
correlators normalized as on the RHS of \eqref{normalize_tensor}, \eqref{normalize_tensor_graviton} and \eqref{normalize_graviton}.

\section{Testing the four-graviton formula}\label{appendix_testingformula}

To test our formula for $\left < \sigma_{p_1} \sigma_{p_2} \sigma_{p_3} \sigma_{p_4} \right >$, we 
have used the results in \cite{Behan:2024srh} for $p_1 = p_2, p_3 = p_4$ and generalized the bootstrap approach in \cite{Behan:2024srh} to compute many new correlators 
for all KK levels. This generalization is straightforward and in the following we will review the key aspects of the method.

\subsection*{Conventions}

We will work at fixed charges $\p$, so let us recall that  
$\sum_{\{\check{s}_{ij}\}}$ becomes a finite sum due to the Gamma functions 
in the denominator. In this case, it is useful to extract a kinematic prefactor in order to write the correlator as a sum 
over finitely many distinct powers of $y$ and $\bar{y}$. When the constraints read 
$\sum_{j = 1}^4 \check{s}_{ij} = p_i - \delta_{\text{S}}$, the number of powers is $\mathcal{E} - \delta_{\text{S}} + 1$ 
where we have defined the \textit{extremality}
\begin{align}
\mathcal{E} = \text{min}\!\left [ \text{min}(p_i), \frac{1}{2} \sum_{i = 1}^4 p_i - \text{max}(p_i) \right ].
\end{align}
In formulae, we parametrize the full correlator as follows,
\begin{align}
\left < \sigma_{p_1} \sigma_{p_2} \sigma_{p_3} \sigma_{p_4} \right > = \textbf{K}\, {C}_{\p}(x,\bar{x},y,\bar{y}) 
\end{align}
where the required transformations under bosonic symmetries are accounted for by
\begin{align}
\textbf{K} = \frac{\left | y_{34}/x_{34} \right |^{p_3+p_4-p_1-p_2 } \left | y_{24}/x_{24} \right |^{p_{2}+p_4-p_1 - p_{3}}\left | y_{23}/x_{23} \right |^{ p_2 + p_{3}-p_1-p_4} }{
\left | y_{14}/x_{14} \right |^{ - 2p_1}\left | y_{23}/x_{23} \right |^{ - 2p_1} } \times 
\frac{  \left | y_{12}/x_{12} \right |^{2\mathcal{E}} \left | y_{34}/x_{34} \right |^{2\mathcal{E}} }{ \left | y_{14}/x_{14} \right |^{2\mathcal{E}}  \left | y_{23}/x_{23} \right |^{2\mathcal{E} } }\,.
\end{align}
Then, we introduce
\begin{equation}
{C}_{\p}= \left [ G^+_{\p} + (\bar{x}^{-1} - x^{-1}) G^-_{\p} \right ]
\end{equation}
where $G^+_{\p}$ has degree $\mathcal{E}$ in in $y^{-1}$ and $\bar{y}^{-1}$, and $G^-_{\p}$ has degree $\mathcal{E} - 1$. Both $G^{\pm}_{\p}$ are also functions of
\begin{equation}
U = x\bar{x}\qquad;\qquad V = (1 - x)(1 - \bar{x}).
\end{equation} 
In our frame, the \textit{superconformal Ward identities} encoding the constraints of fermonic symmetries become simply
\begin{align}
(x \partial_x + y \partial_y) \left [ G^+_{\p} + (\bar{x}^{-1} - x^{-1}) G^-_{\p} \right ] \bigl |_{y = x} = 0 \label{scwi}
\end{align}
and its complex conjugate \cite{Rastelli:2019gtj}. Note that here we will not restrict ourselves to the 
reduced correlator. Rather we will consider the full ${C}_{\p}$, including all contributions that appear 
in the superconformal decomposition, thus both $k,f,\bar{f}$ and ${\cal H}^{\pm}$.

It is useful to write the half-BPS block in this convention. From \cite{Behan:2021}, following \cite[(2.12)]{Behan:2024srh}
\begin{align}
{C}_{\frac{1}{2}\textrm{-BPS}} & = {\bf K}  \left[ \left( \frac{x}{y}\right)^{\!\!\frac{p_{34}}{2} } \left(  \frac{\frac{1-y}{y} }{\frac{ 1-x}{x} }\right)^{\!\!{\cal E} -\frac{ p_1+p_2+p_{34} }{2} }\!\!{\cal G}_h^{\frac{p_{12}}{2},\frac{p_{34}}{2}}(x,y) \times \emph{complex\ conjugate}\right] \\[.2cm]
& = \left | \frac{y_{12}}{x_{12}} \right |^{ \frac{p_1+p_2}{2} }  \left | \frac{y_{34}}{x_{34}} \right |^{ \frac{p_3+p_4}{2} }  \left | \frac{y_{13}}{x_{13}} \right |^{\frac{p_3-p_4}{2} }  \left | \frac{y_{14}}{x_{14}} \right |^{\frac{p_1+p_4-p_2-p_3}{2}}  \left | \frac{y_{24}}{x_{24}} \right |^{\frac{p_2-p_1}{2}}\!\!\!\times\bigg[{\cal G}_h^{\frac{p_{12}}{2},\frac{p_{34}}{2}}(x,y) \times\emph{complex\ conjugate} \bigg]
\end{align} 
where
\begin{equation}\label{connorhalfBPS1}
{\cal G}_h^{\frac{p_{12}}{2},\frac{p_{34}}{2}}(x,y)= g_h^{\frac{p_{12}}{2},\frac{p_{34}}{2}}(x)g_{-h}^{\frac{p_{21}}{2},\frac{p_{43}}{2}}(y) + \lambda_h^{\frac{p_{12}}{2},\frac{p_{34}}{2}}g^{\frac{p_{12}}{2},\frac{p_{34}}{2}}_{1+h}(x) g_{1-h}^{\frac{p_{21}}{2},\frac{p_{43}}{2}}(y) 
\end{equation}
with $g_h^{a,b}(z)=z^h \,_2F_1(h-a,h+b,2h,z)$ the SL(2;$\mathbb{R}$) block, and 
the coefficient $\lambda_h^{a,b}$ given by the formula
\begin{equation}\label{connorhalfBPS2}
 \lambda_h^{a, b} = \frac{(h^2 - a^2)(h^2 - b^2)}{4h^2 (1 - 4h^2)}\,.
\end{equation}
Note that this half-BPS block (together with semishort and long blocks) was given also in \cite{Aprile:2021mvq} using a different notation 
(based on the analytic superspace formalism of \cite{Doobary:2015gia}),
\begin{equation}\label{FAhalfBPS}
{\cal G}_h^{\frac{p_{12}}{2},\frac{p_{34}}{2}}(x,y)= \left(\frac{x}{y}\right)^{\!\!h_{\min}} + \frac{(x-y)}{xy}\sum_{k=0}^{h-h_{\min}-1} g_{h-k}^{\frac{p_{12}}{2},\frac{p_{34}}{2}}(x) g_{k-h+1}^{\frac{p_{21}}{2},\frac{p_{43}}{2}}(y)\qquad;\qquad h_{\min}=\max\left(\frac{p_{12}}{2},\frac{p_{43}}{2}\right)
\end{equation}
Let us recall that the allowed values of $h$ are $h=h_{\min},h_{\min}+1,\ldots, \min(\frac{p_1+p_2}{2},\frac{p_3+p_4}{2})$. 
Then it can be shown that the two expressions \eqref{connorhalfBPS1}-\eqref{connorhalfBPS2} and \eqref{FAhalfBPS} match perfectly.

\subsection*{Details on the bootstrap}

The basic idea of the bootstrap approach  \cite{Behan:2024srh} is to impose the superconformal Ward identities \eqref{scwi} 
and its complex conjugate on the Mellin representations of the full correlator $C_{\p}$, where we parametrize
\begin{align}
G^\pm_{\p} = \int_{-i\infty}^{i\infty} \frac{ds dt}{(4\pi i)^2} U^{\frac{s + 2\mathcal{E} - p_1 - p_2}{2}} V^{\frac{t - 2\mathcal{E} + p_{14}}{2}} \widehat{\mathcal{M}}^\pm \prod_{i < j} \Gamma[-\hat{s}_{ij}]
\end{align}
where $\sum_{j = 1}^4 \hat{s}_{ij} = -p_i$ and
\begin{align}
s = 2\hat{s}_{12} + p_1 + p_2 \qquad;\qquad t = 2\hat{s}_{23} + p_2 + p_3
\end{align}
have been singled out as the independent integration variables. Translating position space differential equations such as \eqref{scwi} into Mellin space difference equations 
can be done with the approach described in \cite{Zhou:2017} (see \cite{Virally:2025} for more recent developments). For these equations to have a unique solution, it is 
important to use an ansatz which includes sufficiently many details about the AdS$_3 \times$S$^3$ effective action. We therefore proceed by inputting the singular parts 
of $\widehat{\mathcal{M}}^\pm$ and using the superconformal Ward identity to bootstrap the regular parts as output. 

A suitable ansatz for the regular part in the plus sector is
\begin{align}
\widehat{\mathcal{M}}^+ \bigl |_{\text{reg}} = \sum_{i, j = 0}^{\mathcal{E}} (b^{i,j}_0 + b^{i,j}_s s + b^{i,j}_t t) y^{-i} \bar{y}^{-j}
\end{align}
where the degree in $s$ and $t$ is fixed by the flat-space limit of supergravity. For the minus sector, 
we must remember that the crossed versions of $\bar{x}^{-1} - x^{-1}$ are
\begin{align}
(x^{-1} - \bar{x}^{-1})U/V \qquad;\qquad (x^{-1} - \bar{x}^{-1})U\,.
\end{align}
This leads to the slightly more involved ansatz
\begin{align}
& \widehat{\mathcal{M}}^- \bigl |_{\text{reg}} =  \sum_{i, j = 0}^{\mathcal{E} - 1}  \frac{(b^{i,j}_{0s} + b^{i,j}_{ss} s + b^{i,j}_{ts} t) }{y^{i} \bar{y}^{j}} 
- \zeta_t \sum_{i, j = 0}^{\mathcal{E} - 1}  \frac{(b^{i,j}_{0t} + b^{i,j}_{st} s + b^{i,j}_{tt} t) }{(1 - y)^{i} (1 - \bar{y})^{j} } 
- \zeta_u \sum_{i, j = 0}^{\mathcal{E} - 1} (b^{i,j}_{0u} + b^{i,j}_{su} s + b^{i,j}_{tu} t) y^i \bar{y}^j.
\end{align}
where
\begin{align}
&
\zeta_t = \frac{(p_1 + p_2 - s)(p_3 + p_4 - s)}{(p_2 + p_3 - t - 2)(p_1 + p_4 - t - 2)} \qquad;\qquad
\zeta_u = \frac{(p_1 + p_2 - s)(p_3 + p_4 - s)}{(2 + p_1 + p_3 - s - t)(2 + p_2 + p_4 - s - t)}.
\end{align}

The singular parts have the expressions
\begin{align}
\widehat{\mathcal{M}}^+ \bigl |_{\text{sing}} &
= \sum_{p = \text{max}(|p_{12}|, |p_{34}|)}^{\text{min}(p_1 + p_2, p_3 + p_4)} C^{\sigma\sigma V}_{p_1p_2p} C^{V \sigma\sigma}_{p p_3p_4} 
\Bigg[  \frac{1}{2} (g^{p_i}_p(y) g^{p_i}_{p - 2}(\bar{y}) + g^{p_i}_{p - 2}(y) g^{p_i}_p(\bar{y})) \mathcal{M}^{2d}_{p,1}  \\
&\rule{5cm}{0pt}+ \lambda^{\frac{p_{12}}{2}, \frac{p_{34}}{2}}_{\frac{p + 1}{2}} g^{p_i}_{p - 1}(y) g^{p_i}_{p - 1}(\bar{y}) \mathcal{M}^{2d}_{p+1, 2}  \nonumber \\
&\rule{5cm}{0pt}+ \lambda^{\frac{p_{12}}{2}, \frac{p_{34}}{2}}_{\frac{p - 1}{2}} (g^{p_i}_{p + 1}(y) g^{p_i}_{p - 3}(\bar{y}) + g^{p_i}_{p - 3}(y) g^{p_i}_{p + 1}(\bar{y})) \mathcal{M}^{2d}_{p+1,0} \nonumber \\
&\rule{5cm}{0pt}+ \frac{1}{2} \lambda^{\frac{p_{12}}{2}, \frac{p_{34}}{2}}_{\frac{p + 1}{2}} \lambda^{\frac{p_{12}}{2}, \frac{p_{34}}{2}}_{\frac{p - 1}{2}} (g^{p_i}_{p - 1}(y) g^{p_i}_{p - 3}(\bar{y}) + g^{p_i}_{p - 3}(y) g^{p_i}_{p - 1}(\bar{y})) \mathcal{M}^{2d}_{p+3,1} \Bigg] \nonumber \\
&+ \sum_{p = \text{max}(|p_{12}|, |p_{34}|)}^{\text{min}(p_1 + p_2, p_3 + p_4) - 2} \!\!\!\!\!\!C^{\sigma\sigma\sigma}_{p_1p_2p} C^{\sigma\sigma\sigma}_{p p_3p_4}  \bigg[ g^{p_i}_p(y) g^{p_i}_p(\bar{y}) \mathcal{M}^{2d}_{p,0} + \frac{1}{2} \lambda^{\frac{p_{12}}{2}, \frac{p_{34}}{2}}_{\frac{p}{2}}(g^{p_i}_p(y) g^{p_i}_{p - 2}(\bar{y}) + g^{p_i}_{p - 2}(y) g^{p_i}_p(\bar{y})) \mathcal{M}^{2d}_{p + 1, 1} \nonumber \\
&\rule{5cm}{0pt}+ \left | \lambda^{\frac{p_{12}}{2}, \frac{p_{34}}{2}}_{\frac{p}{2}} \right |^2 g^{p_i}_{p - 2}(y) g^{p_i}_{p - 2}(\bar{y}) \mathcal{M}^{2d}_{p + 2,0} \Bigg] + \emph{crossing}\\
\widehat{\mathcal{M}}^- \bigl |_{\text{sing}} &= \sum_{p = \text{max}(|p_{12}|, |p_{34}|)}^{\text{min}(p_1 + p_2, p_3 + p_4)} C^{\sigma\sigma V}_{p_1p_2p} C^{V \sigma\sigma}_{p p_3p_4}  \bigg[ 
								\frac{1}{2} (g^{p_i}_p(y) g^{p_i}_{p - 2}(\bar{y}) - g^{p_i}_{p - 2}(y) g^{p_i}_p(\bar{y})) \mathcal{M}^{4d}_{p+1,0}  \nonumber \\
&\rule{5cm}{0pt} + \frac{1}{2} \lambda^{\frac{p_{12}}{2}, \frac{p_{34}}{2}}_{\frac{p + 1}{2}} \lambda^{\frac{p_{12}}{2}, \frac{p_{34}}{2}}_{\frac{p - 1}{2}} (g^{p_i}_{p - 2}(y) g^{p_i}_{p - 4}(\bar{y}) - g^{p_i}_{p - 4}(y) g^{p_i}_{p - 2}(\bar{y})) \mathcal{M}^{4d}_{p+3,0}\bigg] \nonumber \\
&- \sum_{p = \text{max}(|p_{12}|, |p_{34}|)}^{\text{min}(p_1 + p_2, p_3 + p_4) - 2}\!\!\!\!\!\!\!C^{\sigma\sigma\sigma}_{p_1p_2p} C^{\sigma\sigma\sigma}_{p p_3p_4} \left [ \frac{1}{2} \lambda^{\frac{p_{12}}{2}, \frac{p_{34}}{2}}_{\frac{p}{2}} (g^{p_i}_p(y) g^{p_i}_{p - 2}(\bar{y}) - g^{p_i}_{p - 2}(y) g^{p_i}_p(\bar{y})) \mathcal{M}^{4d}_{p+2,0} \right ] + \emph{crossing}
\end{align}
where $\lambda_{h}^{a,b}$ is given in \eqref{connorhalfBPS2}, and the OPE coefficients are
\begin{align}
C^{\sigma\sigma V}_{p_1 p_2 p_3} &=i \frac{p_1^2 + p_2^2 - p_3^2 - 1}{2} \sqrt{\frac{p_1 p_2}{(p_1^2 - 1)(p_2^2 - 1) p_3}} \label{CsigmasigmaV} \\
C^{\sigma\sigma\sigma}_{p_1 p_2 p_3} &= (p_1^2 + p_2^2 + p_3^2 - 2) \sqrt{\frac{p_1 p_2 p_3}{2(p_1^2 - 1)(p_2^2 - 1)(p_3^2 - 1)}}.
\end{align}
These OPE coefficients are determined in \cite{Behan:2024srh}, following the supergravity results in \cite{Arutyunov:2000}.
The factor of $i$ in \eqref{CsigmasigmaV} is justified as in \cite[(3.40)-(3.41)]{Behan:2024srh}.
Above we have used the harmonic polynomials
\begin{align}
g^{p_i}_p(y) &= y^{-\frac{p + p_{34}}{2}} \left(1 - \frac{1}{y}\right)^{\mathcal{E} + \frac{p_{12} - p_{34}}{2}} \times {}_2F_1 \left ( -\frac{p - p_{12}}{2}, -\frac{p + p_{34}}{2}; -p; y \right )
\end{align}
and the Mellin amplitudes for exchange Witten diagrams given by
\begin{align}
\mathcal{M}_{\Delta,\ell} &= \sum_{m = 0}^\infty \frac{f^m_{\Delta, \ell} Q_{\Delta, \ell}(t)}{s - \Delta + \ell + 2m}
\end{align}
where
\begin{align}
f^m_{\Delta, \ell} = \frac{-2^{2 - 2\ell} \Gamma[\Delta + \ell] \Gamma[\tfrac{\Delta + \ell \pm p_{12}}{2}]^{-1} \Gamma[\tfrac{\Delta + \ell \pm p_{34}}{2}]^{-1}}{m! (\Delta - \tfrac{d - 2}{2})_m \Gamma[\tfrac{p_1 + p_2 - \Delta + \ell}{2} - m] \Gamma[\tfrac{p_3 + p_4 - \Delta + \ell}{2} - m]}
\end{align}
serves to truncate the sum. The \textit{Mack polynomial} $Q_{\Delta, \ell}(t)$ has degree 
$\ell$ and explicit expressions in the conventions we have used may be 
found in the appendix of \cite{Behan:2024srh}.

\section{The mixed tensor-graviton correlators}\label{appendix_6Dorigin}

The four-point tensor-graviton correlator was obtained in \cite{Giusto:2020neo} by using a 6D argument, 
and then \cite{Behan:2024srh} demonstrated that this derivation agrees with results obtained using rigorous SCFT$_2$ bootstrap techniques. 
While it was  already known from \cite{Rastelli:2019gtj,Giusto:2019pxc} that the tensor multiplet 
KK tower, $s^i_p(x,y)$, originates from a scalar field $\phi(\z)$ in 6D (of dimension two), 
the authors of \cite{Giusto:2020neo} made the crucial observation that the graviton multiplet 
KK tower, $\sigma_p(x,y)$, originates from a self-dual 3-form field $w_{\mu\nu\rho}(\z)$ 
in 6D (of dimension three) with polarizations appropriately projected. Following \cite{Giusto:2020neo} we shall call this projection $w_{}^{(3)}(\z)$.
As a result, the quest for the four-point tensor-graviton correlators starts from finding the most 
general parametrization of
\begin{equation}
\langle \phi(\z_1)\,\phi(\z_2)\,w^{(3)}\!(\z_3)\,w^{(3)}\!(\z_4)\rangle
\end{equation}
that is compatible with 6D conformal symmetry.

We use here a slightly different notation 
compared to formula \cite[(4.1)]{Giusto:2020neo}.  Our notation is as follows.
We embed 6D coordinates in projective space, 
\begin{equation}
\z_i^M=(X_i^\mu ,Y_i^{a})\qquad;\qquad  X^{\mu}_i=\frac{1}{2} \big( 1\,, x_i\big)\cdot \,\sigma^{\mu}\!\cdot \Big(\,^{\,1}_{\,\bar{x}_i} \Big) \qquad;\qquad Y^{a}_i= \frac{1}{2} \big( 1\,, y_i\big)\cdot \,\sigma_{e}^{a}\!\cdot \Big(\,^{\,1}_{\,\bar{y}_i} \Big) \,,
\end{equation} 
 where $\z_i^{M=1,\ldots 8}$ are 8d projective null coordinates,  $\sigma_e^{a}= (\mathbf{1},-i\vec{\sigma})$, $\sigma^{\mu}= (\mathbf{1},\vec{\sigma})$ with $\vec{\sigma}=\tt{PauliMatrix}$ in \emph{Mathematica}, and distances are measured as:
\begin{equation}
\z_{ij}^2
= \sum_{\mu,\nu} \eta_{\mu\nu}(X_i - X_j)^{\mu}(X_i - X_j)^{\nu} +  \sum_{a,b}\delta_{ab}(Y_i- Y_j)^a(Y_i- Y_j) ^b  = x_{ij}\bar{x}_{ij}-y_{ij}\bar{y}_{ij} \equiv x_{ij}^2-y_{ij}^2 \,,
\end{equation}  
where $\eta_{\mu\nu}=\tt{diagonal}(-1,1,1,1)$.

It turns out that \cite{Giusto:2020neo} 
\begin{equation}\label{RS41}
\langle \phi(\z_1)\,\phi(\z_2)\,w^{(3)}\!(\z_3)\,w^{(3)}\!(\z_4)\rangle=
D_{sss}(\z) {\cal S}_+^{} + D_{[12]}(\z){\cal S}_+^{[12]} + D_{stu}(\z) {\cal T}^{[12]}
\end{equation}
with the $D_i(\z)$ determined aposteriori, and
\begin{align}
\label{Ap}
{\cal S}_+^{} =&\, -\z_{12}^4\, y_{34}^2 x_{34}^2  \\[.2cm]
 \label{App}
{\cal S}_+^{[12]} =&\, +\z_{12}^2\,(y_{14}^2 y_{23}^2 - y_{13}^2 y_{24}^2) { x_{34}^2 } - (x\leftrightarrow y)\\[.2cm]
 \label{Bp}
{\cal T}_+^{[12]} =&\, +\z_{12}^2\frac{ (  y_{14}^2 y_{23}^2 +y_{13}^2 y_{24}^2 ) x_{34}^2 + (x\leftrightarrow y ) }{2} - \Big[ { y_{13}^2 y_{14}^2 x_{23}^2 x_{24}^2 + (x\leftrightarrow y) }{}\Big]  \\
 &
 \label{Bpp}
 - \Big[ y_{12}^2 y_{34}^2 (  - x_{12}^2 x_{34}^2 +x_{13}^2 x_{24}^2+ {x_{14}^2 x_{23}^2 })  
+ y_{13}^2 y_{24}^2( - x_{13}^2 x_{24}^2 +x_{12}^2 x_{34}^2)+ y_{14}^2 y_{23}^2(  - x_{14}^2 x_{23}^2+x_{12}^2 x_{34}^2 )  \Big] \\[.2cm]
{\cal T}_{-}^{[12]}=&\,  +4 \epsilon_{\mu_1\mu_2\mu_3\mu_4} {Y_1^{\mu_1} Y_2^{\mu_2} Y_3^{\mu_3} Y_{4}^{\mu_4} } (x-\bar{x}){x_{13}^2 x_{24}^2}{} 
\end{align}
where we split ${\cal T}^{[12]}={\cal T}_+^{[12]} + {\cal T}_-^{[12]}$ for later convenience. 
After going through definitions in \cite{Giusto:2020neo}, the functions $D_{i}(\z)$ are
\begin{align}
\label{Duno}
D_{sss}(\z)&= \oint\!\frac{d\S d\T}{(2\pi i)^2} \frac{ {\bold U}^{s+4} {\bold V}^{\T} }{(\z_{12}^2 \z_{34}^2)^4} \Gamma[-\S]^2\Gamma[-\T]^2\Gamma[\S+\T+4]^2 \Bigg[  \frac{ 1 }{(\S+1)(\S+2)(\S+3)}\Bigg] \\
\label{Ddue}
D_{[12]}(\z)&= \oint\!\frac{d\S d\T}{(2\pi i)^2}  \frac{ {\bold U}^{\S+4} {\bold V}^{\T} }{(\z_{12}^2 \z_{34}^2)^4}\Gamma[-\S]^2\Gamma[-\T]^2\Gamma[\S+\T+4]^2 \Bigg[  \frac{ \frac{1}{2} }{(\S+1)(\S+2)(\T+1)}+ \frac{ \frac{1}{2} }{(\S+1)(\S+2)(\S+\T+3)}\Bigg] \\
\label{Dtre}
D_{stu}(\z)&= \oint\!\frac{d\S d\T}{(2\pi i)^2}  \frac{ {\bold U}^{\S+4} {\bold V}^{\T} }{(\z_{12}^2 \z_{34}^2)^4} \Gamma[-\S]^2\Gamma[-\T]^2\Gamma[\S+\T+4]^2 \Bigg[\frac{1}{(\S+1)(\T+1)(\S+\T+3)}\Bigg]
\end{align}
where
\begin{equation}
{\bold U}=\frac{\z_{12}^2 \z_{34}^2}{\z_{13}^2 \z_{24}^2 }\qquad;\qquad {\bold V}=\frac{\z_{14}^2 \z_{23}^2 }{\z_{13}^2 \z_{24}^2}\,.
\end{equation}
In the rest of this section we will show how to go from \eqref{RS41} 
to our formula \eqref{tensgravcorr} in the main text.

First of all, the Gram determinant formula gives
\begin{equation}
16 (\epsilon_{\mu \nu \rho \sigma}Y_1^{\mu}Y_2^{\nu}Y_3^{\rho}Y_4^{\sigma})^2 = 16 \det (Y_i \cdot Y_j) = y_{13}^4 y_{24}^4(y {-} \bar{y})^2 \,,
\end{equation}
where in the second equality we used the scalar product $y_{ij}^2=+2\,Y_i \cdot Y_j$. 
Therefore it is immediate to see that ${\cal T}_{-}^{[12]}$ precisely accounts for the prefactor in the 
minus-type correlator \eqref{WardIdentity3}. Together with the function $D_{stu}(\z)$,  they 
correctly reproduce the minus-sector correlator, namely
\begin{equation}
{\cal T}_{-}^{[12]}D_{stu}(\z)\Bigg|_{\p}=-(x {-} \bar{x}) (y{-} \bar{y})\ x_{13}^2 x_{24}^2 y_{13}^2 y_{24}^2 
\!\!\!\!\intsumdue \ \mathcal{M}_{stu}
\end{equation}
which coincides with ${\cal H}^-_{\p}$ given in \eqref{tensgravmin}-\eqref{tensgravMellinmin}.

The rest of the terms  \eqref{Ap}-\eqref{Bpp} only contribute to the plus sector. It will be 
convenient to rearrange them slightly, by noting two things. First, the term $\z_{12}^2\frac{(\ldots)}{2}$ 
of ${\cal T}_+$ wants to go with ${\cal S}_+$ in \eqref{App}. Second, the line \eqref{Bpp} wants to 
have $y^2_{12}y_{34}^2(-x^2_{12} x_{34}^2+ x_{13}^2 x_{24}^2 +x^2_{14}x^2_{23})+\emph{crossing}$, 
which is something that we can achieve by adding and subtracting terms.  
After this minor revision, the correlator in the plus sector reads
\begin{equation}
\langle \phi(\z_1)\,\phi(\z_2)\,w^{(3)}\!(\z_3)\,w^{(3)}\!(\z_4)\rangle^+=
D_{sss}(\z) {\cal S}_+ + \bigg[ D_{sst}(\z){\cal A}_+ + D_{sst}(\z'){\cal A}'_+\bigg] + D_{stu}(\z) {\cal B}_+
\end{equation}
where
\begin{equation}
{\cal A}_+ =+\z_{12}^2( y_{14}^2 y_{23}^2 x_{34}^2 - x_{14}^2 x_{23}^2 y_{34}^2 )\qquad\qquad;\qquad\qquad 
{\cal A}'_+ = \z_{12}^2( y_{13}^2 y_{24}^2 x_{34}^2 - x_{13}^2 x_{24}^2 y_{34}^2 ) \\[.2cm]
\end{equation}
with ${\cal A}'_+$ related to ${\cal A}_+$ by exchanging points $1\leftrightarrow 2$ (or $3\leftrightarrow 4$), corresponding to $\T\leftrightarrow -\S-\T-4$, and
\begin{align}
D_{sst}(\z)=&\oint\!\frac{d\S d\T}{(2\pi i)^2} \frac{ {\bold U}^{\S+4} {\bold V}^{\T} }{(\z_{12}^2 \z_{34}^2)^4} \Gamma[-\S]^2\Gamma[-\T]^2\Gamma[\S+\T+4]^2 \Bigg[ \frac{1 }{(\S+1)(\S+2)(\T+1)} \Bigg]\,, \\
D_{ssu}(\z)=D_{sst}(\z')=&\oint\!\frac{d\S d\T}{(2\pi i)^2} \frac{ {\bold U}^{\S+4} {\bold V}^{\T} }{(\z_{12}^2 \z_{34}^2)^4} \Gamma[-\S]^2\Gamma[-\T]^2\Gamma[\S+\T+4]^2 \Bigg[ \frac{1 }{(\S+1)(\S+2)({\U}+1)} \Bigg]\,.
\end{align}
Finally,\\
\begin{align}
\label{Bplus}
\!\!\!\!{\cal B}_+ =&\,  -  y_{13}^2 y_{14}^2 x_{23}^2 x_{24}^2 - y_{23}^2 y_{24}^2 x_{13}^2 x_{14}^2 + y_{13}^2 y_{24}^2 x_{14}^2 x_{23}^2  + y_{14}^2 y_{23}^2 x_{13}^2 x_{24}^2 \\
%
&- \Big[ y_{12}^2 y_{34}^2 (  - x_{12}^2 x_{34}^2 +x_{13}^2 x_{24}^2+ x_{14}^2 x_{23}^2 ) 
+{ y_{13}^2 y_{24}^2( - x_{13}^2 x_{24}^2 +x_{12}^2 x_{34}^2 + x_{14}^2 x_{23}^2)+ y_{14}^2 y_{23}^2(  - x_{14}^2 x_{23}^2+x_{12}^2 x_{34}^2  +x_{13}^2 x_{24}^2) }{ } \Big] 
\notag
\end{align}

Now, the crucial idea is to rewrite \eqref{Ap}-\eqref{Bpp} as polynomials in 
\begin{equation}
\z_{ab}^2\z_{cd}^2\qquad;\qquad y_{ij}^2\qquad;\qquad G_{ij}\equiv \frac{ y_{ij}^2}{\z_{ij}^2}
\end{equation}
Here $y^2_{ij}$ carry information on the three-form and $G_{ij}$ is the natural ``propagator"  in the sense 
that it also carries the additional information about polarization  vectors (that are not present in the case 
of four external scalars).  Moreover, $G_{ij}={y_{ij}^2}/{x_{ij}^2} +\ldots $ in the S$^3$ expansion, namely, 
the leading term is precisely the AdS$_3\times$S$^3$ superpropagator. In this rewriting we will think 
about $G_{ij}$ as the action  of an operator on the ${D}_{i}(\z)$ functions. The only other operators 
that will act on the ${D}_{i}(\z)$ are the pairs $\z^2_{12}\z^2_{34}$, $\z^2_{14}\z^2_{23}$,  
$\z^2_{13}\z^2_{24}$, corresponding to the $\S$, $\T$, or ${\U}$-channel, or equivalently 6D 
cross ratios up to a common prefactor, as usual in Mellin space.

The rewriting we are looking for can be reconstructed by first replacing $x_{ij}^2$ using $x_{ij}^2= \z_{ij}^2 + y_{ij}^2$
in formulae \eqref{Ap}-\eqref{Bpp}. Then, it will be useful to collect powers of $\z$ under the bookkeeping 
rescaling $\z\rightarrow \sqrt{c}\, \z$. The powers of $c^i$ that appear are $i=0,1,2,3$. 
Let us consider  ${\cal B}_+$ first. Note that of all the terms in ${\cal B}_+$, the first line 
in \eqref{Bplus} does not contribute to $c^0$ or $c^1$. Focussing on the terms $c^0$, 
i.e.~no powers of $c$, these have top degree in the $y^2_{ij}$, thus cannot depend on $G_{ij}$ 
because $D_{stu}(\z)$ is already balanced in the sense of the identity \eqref{identity_balanced}, 
see the next section. The terms of order $c^1$, with three $y^2_{ij}$ and one $\z^2_{ij}$, are also clear: 
up to an overall factor of $\z_{12}^2 \z_{34}^2$ these can be written in terms of a single $G_{ij}$, e.g.~the 
terms proportional to $G_{12}$ and $G_{34}$ can be recognized immediately. The other terms will be 
products of two $y^2_{ij}$ and two $\z^2_{ij}$. At this point it is quite simple to obtain the result we 
are looking for,
\begin{align}
{\cal B}_+=
&\ y_{13}^4 y_{24}^4 (y-\bar{y}) ^2 +  \label{Bline1} \\[.1cm]
& +\left[+ {\cal Y}_s (G_{12}+G_{34})+{\cal Y}_{t}(G_{14}+G_{23}) \frac{\bold V}{\bold U}+{\cal Y}_u(G_{13}+G_{24}) \frac{1}{\bold U} \right] \z_{12}^2 \z_{34}^2  \label{Bline2}\\
&  +\left[ G_{12} G_{34} \left(1-\frac{1}{ \bold U} -\frac{\bold V}{\bold U}\right)  + G_{14} G_{23}  \left(1 -\frac{1}{\bold V}-\frac{\bold U}{\bold V}\right) \frac{\bold V^2}{\bold U^2}  + G_{13} G_{24}\left(1 -{\bold U}-{\bold V}\right) \frac{1}{\bold U^2} \right] \z_{12}^4 \z_{34}^4  \label{Bline3} \\
&+\bigg[  G_{14} G_{23} +G_{13}G_{24} -G_{23}G_{24} -G_{13}G_{14} \bigg]\!(\z_{14}^2 \z_{23}^2)( \z_{13}^2 \z_{24}^2). 
\label{Bline4} 
\end{align}
Note that all terms in line \eqref{Bline2} are simply the crossings of the first one. The same is true for 
the terms in \eqref{Bline3}. The rewriting  of ${\cal B}_+$ respects manifest crossing symmetry. Then, 
we also find
\begin{align}
{\cal A}_+& = \bigg[ G_{14} G_{23}-G_{23}G_{34} -G_{14}G_{34} -G_{34} \bigg]\! (\z_{14}^2 \z_{23}^2 )(\z_{12}^2 \z_{34}^2)
\label{Bline5}  \\
{\cal S}_+ & = \bigg[ -G^2_{34} -G_{34} \bigg]  \z_{12}^4 \z_{34}^4.
\label{Bline6} 
\end{align}
We can now exhibit the rewriting that leads to our formula \eqref{tensgravcorr} in the main text. 
To do so we will need to use the three identities, called \emph{Identity 1,2,3}, and given in the next section.

\subsection*{Useful identities}

Consider a Mellin amplitude ${\cal M}(\S,\T)$. Let $\delta\in\mathbb{N}$, and define 
the integral $I_{\delta}$ as follows   
\begin{align}
I_{\delta}= & \oint \frac{d\S d\T}{(2\pi i)^2}  ({\bold z}_{12}^{2} {\bold z}_{34}^{2})^{\S} ({\bold z}_{14}^{2} {\bold z}_{23}^{2})^{\T}({\bold z}_{13}^{2} {\bold z}_{24}^{2})^{\U}  \Gamma[-\S]^2 \Gamma[-\T]^2 \Gamma[-{\U}]^2  {\cal M}(\S,\T)  \\
= & \oint \frac{d\S d\T}{(2\pi i)^2} \frac{  {\bold U}^{\S+\delta}\,{\bold V}^{\T} }{(\z^2_{12} \z_{34}^2)^{\delta}} \Gamma[-\S]^2 \Gamma[-\T]^2 \Gamma[-{\U}]^2  {\cal M}(\S,\T) 
\end{align}
where ${\U}=-\S-\T-\delta$. In the second line, we rewrote the integrand in terms of the 
cross ratios, in order to make contact with the notation used in the previous section.

\begin{center}\emph{Identity 1}\end{center} 
Let $a,b,c \in \mathbb{Z}$, and let us consider the operation of multiplying $I_{\delta}$ by $\z_{ij}^2\z_{kl}^2$ factors, namely
\begin{equation}
(\z^2_{12} \z^2_{34})^a (\z^2_{14} \z^2_{23})^b (\z^2_{13} \z^2_{24})^c\, I_{\delta}\,.
\end{equation}
This gives 
\begin{align}
\label{identity_withZ}
(\z^2_{12} \z^2_{34})^a (\z^2_{14} \z^2_{23})^b (\z^2_{13} \z^2_{24})^c\, I_{\delta}=  
\oint \frac{d\S d\T}{(2\pi i)^2} \frac{  {\bold U}^{\S+\delta'}\,{\bold V}^{\T} }{(\z^2_{12}\z_{34}^2)^{\delta'}} \Gamma[-\S]^2 \Gamma[-\T]^2 \Gamma[\S+\T+\delta']^2  {\cal N}(\S,\T)
\end{align}
where
\begin{equation}
\delta'=\delta-a-b-c\qquad;\qquad {\cal N}(\S,\T)=(-a+1+\S)_a^2 \, (-b+1+\T)_b^2  \,(\S+\T+\delta')_c^2 \,{\cal M}(\S-a,\T-b) 
\end{equation}
and $(\#)_a$ is the Pochammer symbol.

\begin{center}\emph{Identity 2}\end{center} 
We are interested in expanding $I_{\delta}$ in the $y_{ij}^2$, 
and projecting onto a correlator labelled by external charges $\p=\{p_1,p_2,p_3,p_4\}$.  
This projection, called $\mathscr{P}_{\p}[I_{\delta}]$ in the following, can be achieved by 
replacing $y_{ij}^2\rightarrow a_i a_j y_{ij}^2$ and then taking a contour integral in the $a_i$, 
as follows
\begin{equation}
\mathscr{P}_{\p}[I_{\delta}]=\oint \prod_i \frac{da_i}{(2\pi i)} { a_1^{-p_1}a_2^{-p_2}a_3^{-p_3} a_4^{-p_4} 
\bigg[ I_{\delta}( {\bf z}_{ij}^2= x_{ij}^2- a_i a_j y_{ij}^2) } \bigg]\,.
\end{equation}
Note that the Mellin integral on the RHS of \eqref{identity_withZ} can expanded in the same way. 

When we replace $y^2_{ij}\rightarrow a_i a_j y_{ij}^2$ we can expand the integrand  
using the well known\,$\,_1F_0$ hypergeometric identity
\begin{equation}\label{1F0id}
(1-\chi)^{\mathscr{E}}= \sum_{i=0}^{\infty} \frac{ ( -\mathscr{E})_i }{i!} (+\chi)^i\,.
\end{equation}
This yields the result
\begin{equation}\label{identity_balanced}
I_{\delta}=\sum_{\underline{\Delta} } 
\,a_1^{\Delta_1} a_2^{\Delta_2}a_3^{\Delta_3} a_4^{\Delta_4} 
\ \sum_{\tilde{s},\tilde{t}}  \oint_{-i\infty}^{+i\infty}\!\!\frac{dsdt}{(2\pi i)^2} \prod_{i<j}   { x_{ij}^{2 \hat{s}_{ij}}} {  y_{ij}^{2 \check{s}_{ij}} } \frac{\Gamma[-\hat{s}_{ij}]}{\Gamma[\check{s}_{ij}{+}1] } {\cal M}({\S},{\T}) 
\end{equation}
where
\begin{equation}\label{KKconstraint}
\sum_{j\neq i} \check{s}_{ij} = \Delta_i \qquad;\qquad \sum_{j\neq i} \hat{s}_{ij}=-\delta- \Delta_i
\end{equation}
and where
\begin{equation}\label{onshellIn}
{\S}=\check{s}_{12}+\hat{s}_{12}=\check{s}_{34}+\hat{s}_{34}\qquad;\qquad {\T}=\check{s}_{14}+\hat{s}_{14}=\check{s}_{23}+\hat{s}_{23} \,.
\end{equation}
At this point, when we apply the projector we obtain
\begin{equation}\label{intermedio_scrP}
\mathscr{P}_{\p}[I_{\delta}]=\sum_{ \underline{\Delta} } \oint \prod_i \frac{da_i }{(2\pi i)} a_i^{\Delta_i-p_i} 
\bigg[ 
 \, \sum_{\tilde{s},\tilde{t}}  \oint_{-i\infty}^{+i\infty}\!\!\frac{dsdt}{(2\pi i)^2} \prod_{i<j}   { x_{ij}^{2 \hat{s}_{ij}}} {  y_{ij}^{2 \check{s}_{ij}} } \frac{\Gamma[-\hat{s}_{ij}]}{\Gamma[\check{s}_{ij}{+}1] } {\cal M}({\S},{\T})  \bigg]
\end{equation}
The contour integral gives the relation
\begin{equation}
-p_i+\Delta_i=-1
\end{equation}
Therefore $\Delta_i=p_i-1$, and we obtain,
\begin{mdframed}
\begin{equation}
I_{\delta}=\sum_{\p}\intsum\, {\cal M}({\S},{\T}) \qquad\qquad\textrm{where}\qquad\qquad \begin{array}{l} \sum_{j} \check{s}_{ij}=+p_i-1 \\[.2cm] \sum_{j} \hat{s}_{ij}=-p_i+1- \delta\end{array}\vspace{2mm}
\end{equation}
\end{mdframed}
{\bf Remark:} Note that the factor of $a_i^{\Delta_i}$ in \eqref{intermedio_scrP} is shifted depending 
on the R-symmetry factor in front of the sub-amplitude under consideration. 
For example, the factor $y_{13}^4 y_{24}^4$ in \eqref{triplepolesidentity} gives $a_i^{\Delta_i+2}$. 
Similarly, the factors ${\cal Y}_{s,t,u}$ in \eqref{doublepolesidentity1} give
$a_i^{\Delta_i+1}$.

\begin{center}\emph{Identity 3}\end{center} 
Finally, let us consider the action of $G_{ij}$ on  $I_{\delta}$. 
We shall focus on $G_{12}I_{\delta}$ for illustration, since the other cases will have a similar derivation. 
Then, let us start with
\begin{equation}
G_{12}I_{\delta}= y_{12}^2 \oint \frac{d\S d\T}{(2\pi i)^2}  ({\bold z}_{12}^{2})^{\S-1} ({\bold z}_{34}^{2})^{\S} ({\bold z}_{14}^{2} {\bold z}_{23}^{2})^{\T}({\bold z}_{13}^{2} {\bold z}_{24}^{2})^{\U}  \Gamma[-\S]^2 \Gamma[-\T]^2 \Gamma[-{\U}]^2  {\cal M}(\S,\T).
\end{equation}
We will proceed as in \eqref{identity_balanced}. We replace $y_{ij}^2\rightarrow a_i a_j y_{ij}^2$ and use \eqref{1F0id}. This yields
\begin{equation}
G_{12} I_{\delta} =   \sum_{\underline{\Delta} \ge 0 } (-1)
a_1^{\Delta_1+1} a_2^{\Delta_2+1}a_3^{\Delta_3} a_4^{\Delta_4} 
\ \sum_{\tilde{s}',\tilde{t}'}  \oint_{-i\infty}^{+i\infty}\!\!\frac{ds'dt'}{(2\pi i)^2} \prod_{i<j}   { x_{ij}^{2 \hat{s}'_{ij}}} {  y_{ij}^{2 \check{s}'_{ij}} } \frac{\Gamma[-\hat{s}'_{ij}]}{\Gamma[\check{s}'_{ij}{+}1] }
\frac{y_{12}^2}{x_{12}^2}\frac{(-\hat{s}'_{12})}{\S} {\cal M}(\S,\T)
\end{equation}
where
\begin{equation}
\sum_{j\neq i} \check{s}'_{ij} = \Delta_i \qquad;\qquad \sum_{j\neq i} \hat{s}'_{ij}=-\delta- \Delta_i
\end{equation}
i.e.~the same on-shell constraints as in \eqref{KKconstraint}. At this point one might note that 
\begin{equation}
\frac{\Gamma[-\hat{s}'_{12} ](-\hat{s}'_{12})}{\Gamma[\check{s}'_{12}+1]}\frac{(y_{12}^2)^{ +\check{s}'_{12}+1} }{ (x_{12}^2)^{-\hat{s}'_{12}+1} } = 
\frac{\Gamma[-\hat{s}_{12}] (\check{s}_{12})}{\Gamma[\check{s}_{12}+1]} y_{12}^{ 2\check{s}_{12}}  x_{12}^{2\hat{s}_{12}}  \qquad\textrm{where}\qquad \begin{array}{c} 
\check{s}_{12}=\check{s}'_{12}+1\\[.2cm]
\hat{s}_{12}=\hat{s}'_{12}-1 
\end{array}.
\end{equation}
Trivially changing the other variables, i.e.~$\hat{s}_{ij}=\hat{s}'_{ij}$ and $\check{s}_{ij}=\check{s}'_{ij}$ for $(ij)\neq(12)$, the final result reads
\begin{equation}\label{G12ident}
G_{12} I_{\delta} =   \sum_{\underline{\Delta} \ge 0 } (-1)
a_1^{\Delta_1+1} a_2^{\Delta_2+1}a_3^{\Delta_3} a_4^{\Delta_4} 
\ \sum_{\tilde{s},\tilde{t}}  \oint_{-i\infty}^{+i\infty}\!\!\frac{dsdt}{(2\pi i)^2} \prod_{i<j}   { x_{ij}^{2 \hat{s}_{ij}}} {  y_{ij}^{2 \check{s}_{ij}} } \frac{\Gamma[-\hat{s}_{ij}]}{\Gamma[\check{s}_{ij}{+}1] } \frac{(\check{s}_{12})}{\S} {\cal M}(\S,\T)
\end{equation}
where
\begin{equation}
\sum_{j\neq i} \check{s}_{ij} = \left[\begin{array}{c}\Delta_1+1\\ \Delta_2+1 \\ \Delta_3 \\ \Delta_4 \end{array}\right] \qquad;\qquad \sum_{j\neq i} \hat{s}_{ij}=-\delta-\left[\begin{array}{c}\Delta_1+1\\ \Delta_2+1 \\ \Delta_3 \\ \Delta_4 \end{array}\right].
\end{equation}
Crucially, ${\bold s}$ remains unchanged.

At this point, when we apply the projector,
\begin{equation}
\mathscr{P}_{\p}[I_{\delta}]=\sum_{ \underline{\Delta} } \oint \prod_i \frac{da_i}{(2\pi i)} a_1^{-p_1}a_2^{-p_2}a_3^{-p_3} a_4^{-p_4} 
\bigg[ 
 (-1)
 \,a_1^{\Delta_1+1} a_2^{\Delta_2+1}a_3^{\Delta_3} a_4^{\Delta_4}  \ldots \bigg]
\end{equation}
 the contour integral localises on 
\begin{equation}
-p_i+\Delta_i+1=-1\Big|_{i=1,2}\qquad;\qquad -p_i+\Delta_i=-1\Big|_{i=3,4}
\end{equation}
Therefore $\Delta_i=p_i-2$ for $i=1,2$ and $\Delta_i=p_i-1$ for $i=3,4$, and we obtain,
\begin{mdframed}
\begin{equation}
G_{12} I_{\delta}=-\sum_{\p}\intsum\,\bigg[ \check{s}_{12}\,\frac{{\cal M}(\S,\T) }{\S} \bigg] \qquad\qquad\textrm{where}\qquad\qquad \begin{array}{l} \sum_{j} \check{s}_{ij}=+p_i-1 \\[.2cm] \sum_{j} \hat{s}_{ij}=-p_i+1- \delta\end{array}\vspace{2mm}
\end{equation}
\end{mdframed}
The general case is 
\begin{mdframed}
\begin{equation}
\,\left( G_{ij}\right)^a I_{\delta}\,
=(-1)^a\sum_{\p}\intsum\,\bigg[ (\check{s}_{ij}-a+1)_a\,\frac{{\cal M}(\S,\T) }{(\hat{s}_{ij}+\check{s}_{ij}-1+a)_a} \bigg] \qquad\quad\textrm{where}\qquad\quad \begin{array}{l} \sum_{j} \check{s}_{ij}=+p_i-1 \\[.2cm] \sum_{j} \hat{s}_{ij}=-p_i+1- \delta\end{array}\vspace{2mm}
\end{equation}
\end{mdframed}
{\bf Remark:} A similar consideration as in \emph{identity 2} applies here. Namely, the factor of $a_i^{\Delta_i}$ in \eqref{intermedio_scrP} is shifted depending on the R-symmetry factor in front of the sub-amplitude under consideration.

\subsection*{From 6D to AdS$_3\times$S$^3$ with pencil-and-paper}

We are now ready to obtain our formula \eqref{tensgravcorr} in the main text, from the 6D result. It 
will be convenient to divide the computation into triple, double and simple poles. 


{\bf Triple Poles.}
This term is immediately obtained from \eqref{Bline1},
\begin{equation}\label{triplepolesidentity}
y_{13}^4 y_{24}^4 (y-\bar{y}) ^2 D_{stu}(\z)\Bigg|_{\p} \ =\!\!\!\!\intsumtre\left( y_{13}^4  y_{24}^4 (y-\bar{y})^2  \mathcal{M}_{stu} \right)
\end{equation}
where we only used the KK reduction of the RHS through \emph{identity 2}. 


{\bf Double Poles.} There will be two contributions. The first one is immediately found from \eqref{Bline2}, namely
\begin{align}\label{doublepolesidentity1}
\bigg[ + {\cal Y}_s (G_{12}+G_{34})+ \emph{crossing}\,\bigg]\z_{12}^2 \z_{34}^2 D_{stu}(\z)\Bigg|_{\p} \ =\!\!\!\!\intsumdue\left( - {\cal Y}_s  \frac{\check{s}_{12} + \check{s}_{34}}{ (\T+1)(\S+\T+2) }+\emph{crossing} \right) 
\end{align}
In order, we first used \emph{identity 1}, bringing the Mellin amplitude to 
${\S}/({(\T+1)(\S+\T+2)})$ with $\delta=3$, and then we used \emph{identity 3} to 
obtain the final result.  Note that one effect of  \emph{identity 1} is to lower the value 
of $\delta$: for $D_{stu}(\z)$ this was initially $\delta=4$, but it becomes $\delta=3$ 
after applying $\z_{12}^2 \z_{34}^2$. In particular, the value of $\U$ is now $\U=-\S-\T-3$, 
thus $\U+1=-\S-\T-2$.  

To complete the double poles we will have to consider 
the next line, \eqref{Bline3}, namely
\begin{align}
&
\!\!\!\!\!\left[ G_{12} G_{34} \left(1-\frac{1}{ \bold U} -\frac{\bold V}{\bold U}\right)  +\emph{crossing}  \right] \z_{12}^4 \z_{34}^4 D_{stu}(\z)\Bigg|_{\p}= \\[.1cm]
&\rule{.3cm}{0pt}\label{doublepolesidentity2}
=\!\!\!\!\intsumuno  \bigg[ \check{s}_{12} \check{s}_{34}\bigg[ \frac{\S-1}{(\T+1)(\S+\T+1)} - \frac{\S+\T+2}{\S(\T+1)} - \frac{\T}{\S(\S+\T+1)} \bigg] + \emph{crossing}\bigg] \\
&\rule{.3cm}{0pt}\label{doublepolesidentity3}
=\!\!\!\!\intsumuno  \bigg[ \check{s}_{12} \check{s}_{34}\bigg[ -\frac{1}{(\T+1)(\S+\T+1) }- \frac{1}{\S(\T+1)} +\frac{1}{\S(\S+\T+1)} -\frac{2}{\S}\bigg] + \emph{crossing}\bigg]\\
&\label{doublepolesidentity4}
\rule{.3cm}{0pt}
= \!\!\!\!\intsumdue \bigg[ -\frac{y^2_{12} y^2_{34} }{(\T+1)(\S+\T+2) }- \frac{y^2_{12} y^2_{34}}{(\S+1)(\T+1)} +\frac{y^2_{12} y^2_{34}}{(\S+1)(\S+\T+2)} + \emph{crossing} \bigg] \ +\!\!\!\!\!\!\!\!\intsumuno\, \bigg[-\frac{2 \check{s}_{12}\check{s}_{34}}{\S} + \emph{crossing}\bigg].
\end{align}
In \eqref{doublepolesidentity2} we first used \emph{identity 1} to absorb the various 
prefactors and then \emph{identity 3}. In \eqref{doublepolesidentity3}
we just rewrote the result. Note that $G_{12} G_{34} \z_{12}^4 \z_{34}^4 D_{stu}(\z)$ is 
balanced in the sense that it can be split into double and single poles with unit coefficients 
in the numerators.  Similarly the other contributions. Finally in \eqref{doublepolesidentity4} 
we absorbed $\check{s}_{12}\check{s}_{34}$ in those amplitudes that contribute to double 
poles. When doing so we also have to shift the on-shell constraints in $\check{s}_{ij}$, and the bold-face 
variable ${\bold s}\rightarrow {\bold s}+1$. Recall that ${\bold s}={\bold s}_{12}=  {\bold s}_{34}$. 
Collecting the double poles  in \eqref{doublepolesidentity1} and \eqref{doublepolesidentity4} 
we obtain ${\cal Y}_s{\cal M}_{tu}+\emph{crossing}$. See our formulae \eqref{tensgravcorr} and  \eqref{Mst}-\eqref{Mtu} in the main text.


{\bf Simple Poles.} There will be in total four contributions to be taken into account. The first one is given already 
in \eqref{doublepolesidentity4}. The others come from \eqref{Bline4}, \eqref{Bline5} and its crossing, and finally \eqref{Bline6}. Proceeding in the same order, 
and with the same technique as above, we find
\begin{align}
&
\!\!\!\!\!\!\!\!\!\bigg[  G_{14} G_{23} +G_{13}G_{24} -G_{23}G_{24} -G_{13}G_{14} \bigg]\!(\z_{14}^2 \z_{23}^2)( \z_{13}^2 \z_{24}^2) D_{stu}(\z)\Bigg|_{\p} 
= \!\!\!\!\intsumuno \bigg[ -\frac{\check{s}_{14}\check{s}_{23} }{\T^2} - \frac{\check{s}_{13}\check{s}_{24} }{\U^2} + \frac{\check{s}_{23}\check{s}_{24}+\check{s}_{13}\check{s}_{14}}{\T \U} \bigg] \frac{\T\U}{(\S+1)} \\
&
\!\!\!\!\!\bigg[ G_{14} G_{23}-G_{23}G_{34} -G_{14}G_{34} -G_{34} \bigg]\! (\z_{14}^2 \z_{23}^2 )(\z_{12}^2 \z_{34}^2)D_{sst}(\z)\Bigg|_{\p} 
\ \ =\!\!\!\!\intsumuno \bigg[ +\frac{\check{s}_{14}\check{s}_{23} }{\T^2} - \frac{\check{s}_{23}\check{s}_{34}+\check{s}_{14}\check{s}_{34}}{\S \T} + \frac{\check{s}_{34}}{\S} \bigg] \frac{\S\T}{(\S+1)}  \\
&
\bigg[ -G^2_{34}-G_{34} \bigg]  \z_{12}^4 \z_{34}^4 D_{sss}(\z)\Bigg|_{\p}\ \ =
\!\!\!\!\intsumuno \bigg[ - \frac{ (\check{s}_{34}-1)\check{s}_{34}}{ (\S-1)\S} + \frac{\check{s}_{34}}{\S} \bigg] \frac{(\S-1)\S}{(\S+1)}\ \ = 
\!\!\!\!\intsumuno \bigg[ - \frac{ \check{s}^2_{34}}{ (\S+1)} + \frac{\check{s}_{34} \S }{(\S+1)} \bigg] 
\end{align}
where $\U=-\S-\T-2$.
By adding all the terms we find 
\begin{align}
\!\!\!\!\intsumuno\ \Bigg[ &-\frac{2 \check{s}_{12}\check{s}_{34}}{\S}+ \frac{\check{s}_{14}\check{s}_{23}}{(\S+1)} \bigg[ -\frac{ 2(\S+1) }{\T}  -\frac{\U}{\T} +\frac{\S}{\T}\bigg] + 
 \frac{\check{s}_{13}\check{s}_{24}}{(\S+1)} \bigg[  -\frac{ 2(\S+1) }{\U} -\frac{\T}{\U} +\frac{\S}{\U}\bigg] + \frac{- \check{s}^2_{34}+\check{s}_{34}( \S+\T+\U)}{(\S+1)}+\notag\\
 &+ \frac{\check{s}_{23}\check{s}_{24}+\check{s}_{13}\check{s}_{14}-\check{s}_{23}\check{s}_{34}-\check{s}_{14}\check{s}_{34}-\check{s}_{13}\check{s}_{34}-\check{s}_{24}\check{s}_{34} }{(\S+1)} \Bigg] = \\
 \intsumuno\ \Bigg[ &-\frac{2 \check{s}_{12}\check{s}_{34}}{\S} + \frac{\check{s}_{14}\check{s}_{23}+\check{s}_{13}\check{s}_{24}+\check{s}_{23}\check{s}_{24}+\check{s}_{13}\check{s}_{14}-\check{s}_{23}\check{s}_{34}-\check{s}_{14}\check{s}_{34}-\check{s}_{13}\check{s}_{34}-\check{s}_{24}\check{s}_{34}}{\S+1}- \frac{\check{s}^2_{34}+2\check{s}_{34}}{\S+1} \Bigg]
\end{align}
By collecting the $1/(\S+1)$ terms in the middle, we use the on-shell constraints to rewrite the numerator as
\begin{equation}
(\check{s}_{13}+\check{s}_{23})(\check{s}_{14}+\check{s}_{24})-(\check{s}_{13}+\check{s}_{23}+\check{s}_{14}+\check{s}_{24})\check{s}_{34}=(\Delta_3-\check{s}_{34})(\Delta_4-\check{s}_{34})- (\Delta_3-\check{s}_{34}+\Delta_4-\check{s}_{34})\check{s}_{34}
\end{equation}
where $\Delta_i=p_i-1$. All together, the final result  is,
\begin{align}
 \intsumuno\ \Bigg[ &-\frac{2 \check{s}_{12}\check{s}_{34}}{\S} +\frac{\Delta_3\Delta_4-2(\Delta_3+\Delta_4+1-\check{s}_{34})\check{s}_{34}}{\S+1} \Bigg]\  =
 \!\!\!\!\! \intsumdue\ \bigg[-\frac{2 {y}^2_{12}{y}^2_{34}}{\S+1}\bigg] +\!\!\!\!\!\!\!\intsumuno\bigg[ +\frac{2f_{34}}{\S+1} \bigg].
\end{align}
Recalling that $f_{ij}= 
\left(\!\frac{p_i+p_j-1}{2}-\check{s}_{ij}\!\right)^{\!\!2} - \frac{p_i^2+p_j^2-1}{4}$ we find a perfect match with formulae \eqref{tensgravplusMellin1} and \eqref{M2stensgrav} in the main text.

\bibliographystyle{apsrev4-2}

\end{document}